%% file: main.tex
\documentclass[conference]{IEEEtran}
\IEEEoverridecommandlockouts
\usepackage[mathscr]{eucal}
\usepackage{amsmath,amssymb,amsfonts,amsthm,bm}
\usepackage{enumitem}
\usepackage{multirow}
\usepackage{cite}
\usepackage{booktabs}
\usepackage{scalerel}
\usepackage[multiple]{footmisc}
\usepackage{tikz}
\usepackage{pgfplots}
\usepackage{pgfplotstable}
\usepgfplotslibrary{statistics}
\usepgfplotslibrary{groupplots}
\usepgfplotslibrary{colormaps}
\pgfplotsset{compat=1.18}
\usepackage{pifont}
\usepackage{algorithm} %,algorithmic
\usepackage{algpseudocode}
\usepackage{tabularx}
\usepackage{footnote}
\usepackage{balance}
\usepackage{graphicx}
\usepackage{subcaption}
\usepackage{tikz}
\usepackage{xcolor}  % 색상 정의
\usepackage{soul} 
\usetikzlibrary{patterns}

\newcommand{\multiline}[1]{%
  \begin{tabularx}{\dimexpr\linewidth-\ALG@thistlm}[t]{@{}X@{}}
    #1
  \end{tabularx}
}

\pgfplotsset{compat=1.18}
\newenvironment{customlegend}[1][]{%
  \begingroup
  \csname pgfplots@init@cleared@structures\endcsname
  \pgfplotsset{#1}%
}{%
  \csname pgfplots@createlegend\endcsname
  \endgroup
}%
\def\addlegendimage{\csname pgfplots@addlegendimage\endcsname}
\usetikzlibrary{patterns}
\usetikzlibrary{pgfplots.groupplots}
\usetikzlibrary{
  pgfplots.colorbrewer,
}
\pgfplotsset{
  cycle list/.define={my marks}{
    every mark/.append style={solid,fill=\pgfkeysvalueof{/pgfplots/mark list fill}},mark=*\\
    every mark/.append style={solid,fill=\pgfkeysvalueof{/pgfplots/mark list fill}},mark=square*\\
    every mark/.append style={solid,fill=\pgfkeysvalueof{/pgfplots/mark list fill}},mark=triangle*\\
    every mark/.append style={solid,fill=\pgfkeysvalueof{/pgfplots/mark list fill}},mark=diamond*\\
  },
}
\newcommand{\xmark}{\ding{55}}
\DeclareMathSymbol{\mh}{\mathord}{operators}{`\-}
\definecolor{aa}{rgb}{0.19,0.55,0.91}
\definecolor{bb}{rgb}{0.9,0.17,0.31}
\definecolor{cc}{rgb}{0.514,0.325,0.831}
\definecolor{dd}{rgb}{0.12, 0.3, 0.17}
\definecolor{ee}{rgb}{1.0, 0.75, 0.0}
\definecolor{ff}{rgb}{0.01, 0.28, 1.0}
\definecolor{gg}{rgb}{0.65, 0.16, 0.16}

\def\BibTeX{{\rm B\kern-.05em{\sc i\kern-.025em b}\kern-.08em
    T\kern-.1667em\lower.7ex\hbox{E}\kern-.125emX}}
\begin{document}

\title{HyBridge: Bridge-Guided Self-Supervised Hyperbolic Hypergraph Representation Learning}

% \author{\IEEEauthorblockN{So-Bin Jung\IEEEauthorrefmark{1}$^1$, David Yoon Suk Kang\IEEEauthorrefmark{1}$^2$\thanks{\IEEEauthorrefmark{1}Two first authors have contributed equally to this work.}, and Sang-Wook Kim\IEEEauthorrefmark{2}$^1$\thanks{\IEEEauthorrefmark{2} Correponding author.}}
% \IEEEauthorblockA{$^1$\textit{Hanyang University}, South Korea, $^2$\textit{Chungbuk National University}, South Korea \\
% sobin98@hanyang.ac.kr, dyskang@cbnu.ac.kr, wook@hanyang.ac.kr}
% }
\author{\IEEEauthorblockN{So-Bin Jung}
\IEEEauthorblockA{\textit{Hanyang University}\\
Seoul, South Korea \\
sobin98@hanyang.ac.kr}
\and
\IEEEauthorblockN{David Yoon Suk Kang}
\IEEEauthorblockA{\textit{Chungbuk National University}\\
Cheongju, South Korea \\
dyskang@cbnu.ac.kr}
\and
\IEEEauthorblockN{Sang-Wook Kim}
\IEEEauthorblockA{\textit{Hanyang University}\\
Seoul, South Korea \\
wook@hanyang.ac.kr}

}

\maketitle

\definecolor{rank1}{HTML}{FFD966} % 1등: 진한 노란색
\definecolor{rank2}{HTML}{C2F0C2} % 2등: 연두색
\newcommand{\hly}[1]{{\sethlcolor{rank1}\hl{#1}}} % 노란색 블록
\newcommand{\hlb}[1]{{\sethlcolor{rank2}\hl{#1}}}   % 녹색/하늘색 블록

\newcommand{\ie}{{\it i.e.}}
\newcommand{\eg}{{\it e.g.}}
\newcommand{\hbc}{{\small\textsf{HyBridge}}}

\begin{abstract}
Hypergraphs capture groupwise relationships among multiple nodes and often exhibit tree-like structures. Although hyperbolic space is well suited for modeling such structures, simply extending existing hypergraph representation learning (HRL) methods to hyperbolic space does not consistently improve performance. In this paper, we investigate the relationship between node structural properties and hyperbolic embedding radii and find that bridge nodes tend to lie closer to the center in graphs, whereas this relationship is not consistently observed in hypergraphs. We further find that conventional bridge score estimation may fail to capture the groupwise connectivity of hypergraphs. Motivated by these findings, we propose \hbc, a self-supervised hyperbolic HRL framework that (1) estimates node bridge scores by considering the external connectivity of individual hyperedges, (2) learns node and hyperedge embeddings through hyperbolic hypergraph encoding, (3) guides node radial positions according to their bridge scores, and (4) employs contrastive learning to preserve node discriminability. Extensive experiments on 10 real-world hypergraphs demonstrate that \hbc\ outperforms 22 competing methods, achieving performance improvements of up to 9.82\% in node classification. For reproducibility, we provide the code at https://anonymous.4open.science/r/HyBridge/.
\end{abstract}

\begin{IEEEkeywords}
Hypergraph representation learning, hyperbolic representation learning, bridge-guided learning, contrastive learning, self-supervised learning
\end{IEEEkeywords}

\input{s1}
\input{s2}
\input{s3}
\input{s4}

\input{s5}
\input{s6}

% \section*{Acknowledgment}

% The preferred spelling of the word ``acknowledgment'' in America is without 
% an ``e'' after the ``g''. Avoid the stilted expression ``one of us (R. B. 
% G.) thanks $\ldots$''. Instead, try ``R. B. G. thanks$\ldots$''. Put sponsor 
% acknowledgments in the unnumbered footnote on the first page.

% \section*{References}

% Please number citations consecutively within brackets \cite{b1}. The 
% sentence punctuation follows the bracket \cite{b2}. Refer simply to the reference 
% number, as in \cite{b3}---do not use ``Ref. \cite{b3}'' or ``reference \cite{b3}'' except at 
% the beginning of a sentence: ``Reference \cite{b3} was the first $\ldots$''

% Number footnotes separately in superscripts. Place the actual footnote at 
% the bottom of the column in which it was cited. Do not put footnotes in the 
% abstract or reference list. Use letters for table footnotes.

% Unless there are six authors or more give all authors' names; do not use 
% ``et al.''. Papers that have not been published, even if they have been 
% submitted for publication, should be cited as ``unpublished'' \cite{b4}. Papers 
% that have been accepted for publication should be cited as ``in press'' \cite{b5}. 
% Capitalize only the first word in a paper title, except for proper nouns and 
% element symbols.

% For papers published in translation journals, please give the English 
% citation first, followed by the original foreign-language citation \cite{b6}.

\bibliographystyle{IEEEtran}
\bibliography{temp.bib}

\end{document}

%% file: s1.tex
\section{Introduction}\label{s1}
\textit{Graph Representation Learning (GRL)} aims to learn node embeddings by capturing graph structures and node attributes for downstream tasks, such as node classification and link prediction~\cite{wan18:aaai,kipf17:iclr,hou23:www,you20:nips,lee20:sigir}.
However, many real-world systems involve \textit{groupwise} relations, where multiple entities jointly participate in a single interaction, as commonly observed in co-authorship networks and biological systems.
Conventional graphs, which represent interactions as \textit{pairwise} relations between nodes, are inherently limited in capturing such higher-order interactions.
In contrast, hypergraphs naturally model \textit{groupwise} relations by allowing each hyperedge to connect more than two nodes.
Accordingly, \textit{Hypergraph Representation Learning (HRL)} aims to learn node embeddings that jointly capture higher-order structural information encoded by hyperedges and node attributes~\cite{yad19:neurips,don20:icmlw,fen19:aaai,zha20:iclr}.

Most existing GRL and HRL methods learn node representations in \textit{Euclidean space}~\cite{wan18:aaai,kipf17:iclr,hou23:www,you20:nips,lee20:sigir,yad19:neurips,don20:icmlw,fen19:aaai,zha20:iclr}.
However, many real-world graphs and hypergraphs exhibit tree-like structures~\cite{liu19:nips,agarwal22:icdm}, which are fundamentally difficult to represent faithfully in Euclidean space.
In a regular tree where each node has $b$ children, the number of nodes at distance $r$ from the root grows \textit{exponentially} as $b^r$, whereas the volume of a $d$-dimensional Euclidean ball of radius $r$ grows only \textit{polynomially} as $r^d$.
This mismatch in growth rates can introduce substantial structural distortion when embedding tree-like structures into Euclidean space~\cite{nickel17:nips,liu19:nips}.

To alleviate these limitations, \textit{Hyperbolic Graph Representation Learning (HB-GRL)} has been studied~\cite{dai21:cvpr,shimizu21:iclr,gulcehre19:iclr,liu19:nips,chami19:nips,jiao24:icde,feng21:icde}.
Hyperbolic space has negative curvature and expands \textit{exponentially} with radius, making it well suited for representing tree-like structures~\cite{liu19:nips,chami19:nips,nickel17:nips}.
Indeed, previous HB-GRL studies have shown that hyperbolic methods tend to perform particularly well on graph datasets with stronger tree-like structures~\cite{chami19:nips,liu19:nips}.
Prior HB-GRL and HB-HRL studies have employed \textit{hyperbolicity}~\cite{gromov87:book} as a measure of tree-likeness, where lower hyperbolicity generally indicates a stronger tree-like structure~\cite{chami19:nips,agarwal22:icdm}.
Motivated by this geometric property, we first examine the hyperbolicity of real-world graph and hypergraph datasets and find that hypergraphs consistently exhibit relatively low hyperbolicity (see Section~\ref{s3.1}), suggesting that hyperbolic space may be well suited for HRL.

Despite this geometric compatibility, HB-HRL has received relatively limited attention compared with HB-GRL~\cite{agarwal22:icdm}.
We therefore extend representative Euclidean HRL models to hyperbolic space and compare their node classification performance with that of their Euclidean counterparts (see Section~\ref{s3.2}).
Surprisingly, the results show that hyperbolic models do not consistently outperform their Euclidean counterparts.
This finding suggests that merely adopting hyperbolic geometry is insufficient and that effectively leveraging its geometric properties may be crucial for HB-HRL.
We therefore draw insights from hyperbolic graph embeddings, where HB-GRL has been extensively studied, to identify structural guidance that can be leveraged for HB-HRL.
To this end, we first examine how nodes are positioned in hyperbolic space for graphs.

In hyperbolic graph embeddings, nodes have been shown to form a \textit{hierarchical center--periphery structure}, with higher-level nodes in a tree-like structure positioned closer to the center and lower-level nodes positioned farther away~\cite{chami19:nips}.
In such a structure, the distance of a node embedding from the origin (\ie, center), referred to as its \textit{radial position}, reflects its position in the underlying hierarchy.
This motivates us to investigate which node structural properties are associated with radial positions in HB-GRL.
Using HB-GCN~\cite{chami19:nips}, we analyze the relationship between node radial positions and various structural properties (see Section~\ref{s3.3}).
Our analysis reveals that nodes closer to the center exhibit more diverse neighborhood labels, a characteristic of \textit{bridge nodes} that connect different node groups~\cite{he16:kdd,platonov23:nips,zhu21:aaai,wang24:icml,hua26:tfs}.
This finding reveals a connection between a node's bridging role and its radial position: nodes with stronger bridging roles tend to be positioned closer to the center.
Moreover, guiding nodes with higher bridge scores toward the center improves node classification performance, demonstrating the effectiveness of bridge scores as structural guidance for hyperbolic node representations.

We next examine whether the same relationship holds in hypergraphs (see Section~\ref{s3.4}).
In contrast to graphs, conventional bridge scores do not exhibit a consistent relationship with radial positions, and bridge-guided learning does not consistently improve node classification performance.
Conventional bridge score estimation relies on the label distribution of each node's direct neighbors~\cite{he16:kdd,hua26:tfs}.
In hypergraphs, however, such node-level aggregation can obscure the distinct bridging roles of individual hyperedges, as nodes within the same hyperedge tend to exhibit highly similar neighborhood label distributions despite participating in different hyperedges (see Section~\ref{s3.5}).
These observations suggest that conventional bridge scores may not adequately capture the groupwise connectivity inherent in hypergraphs, motivating a hypergraph-specific bridge score that explicitly accounts for the distinct bridging roles of individual hyperedges.

To this end, we propose {\hbc}, a bridge-guided HB-HRL framework that estimates node-level bridging roles by explicitly accounting for groupwise connectivity and encourages nodes with stronger bridging roles to be positioned closer to the center of hyperbolic space.
{\hbc} consists of four main steps.
First, without relying on ground-truth node labels, {\hbc} generates pseudo-cluster labels from the hypergraph connectivity structure using a clustering approach adopted in prior HRL studies~\cite{du22:bd}.
It then estimates the bridge score of each hyperedge based on the pseudo-cluster label distribution of nodes in neighboring hyperedges but outside the target hyperedge and obtains each node's bridge score by aggregating the scores of its incident hyperedges.
Next, {\hbc} learns node and hyperedge representations in hyperbolic space using a hyperbolic hypergraph encoder.
Then, {\hbc} incorporates a bridge-guided loss that encourages nodes with higher bridge scores to be positioned closer to the center.
Finally, {\hbc} employs contrastive learning to preserve representation discriminability among nodes with similar bridge scores.

Extensive experiments on 10 real-world hypergraph datasets against 22 state-of-the-art competing methods demonstrate the effectiveness of {\hbc} across node classification and hyperedge prediction, with improvements of up to 9.82\% in node classification.
Furthermore, our ablation studies demonstrate the complementary roles of hyperbolic geometry and bridge-guided learning: hyperbolic representations alone do not consistently outperform their Euclidean counterparts, whereas incorporating bridge guidance consistently improves the effectiveness of hyperbolic representations.

The contributions of this paper are summarized as follows:
\begin{itemize}[leftmargin=*]
    \item \textbf{Important observations.} We investigate the relationship between node structural properties and radial positions in hyperbolic space and identify two key observations.
    \begin{itemize}
        \item \textbf{(O1)} In graphs, nodes with higher bridge scores tend to be positioned closer to the center of hyperbolic space.
        \item \textbf{(O2)} In hypergraphs, this relationship is not consistently observed, and conventional bridge score estimation may not adequately capture hyperedge-level groupwise connectivity.
    \end{itemize}

    \item \textbf{Novel framework.} Based on these observations, we propose {\hbc}, a self-supervised HB-HRL framework.
    \begin{itemize}
        \item \textbf{(C1)} Hypergraph-specific bridge score estimation that captures groupwise connectivity by explicitly accounting for the distinct bridging roles of individual hyperedges.
        \item \textbf{(C2)} Bridge-guided learning that uses the estimated bridge scores as structural guidance for node radial positions, together with contrastive learning to preserve representation discriminability.
    \end{itemize}

    \item \textbf{Extensive evaluation.} We evaluate {\hbc} against 22 state-of-the-art competing methods on 10 real-world hypergraph datasets across node classification and hyperedge prediction tasks.
\end{itemize}

%% file: s2.tex
\section{Related Work}\label{s2}
\noindent\textbf{Representation Learning in Euclidean Space.}  
Existing GRL methods learn node representations in Euclidean space by leveraging graph structures and node features~\cite{wan18:aaai,velickovic18:iclr,kipf17:iclr}.
More recently, self-supervised GRL methods have been developed to learn node representations without node labels through contrastive or generative objectives~\cite{hou23:www, li23:kdd,you20:nips,zhe22:nips}.

To capture groupwise relationships beyond pairwise graph structures, various HRL methods have been proposed~\cite{fen19:aaai, don20:icmlw}.
While early methods transform hypergraph structures into graph-based representations~\cite{fen19:aaai,yad19:neurips}, subsequent approaches directly model groupwise relationships through hyperedge-based message passing~\cite{don20:icmlw,hua21:ijcai,chi21:iclr,wan23:iclr,wan23:icml}.
Recent self-supervised HRL methods learn node representations without node labels through contrastive learning~\cite{lee23:aaai,wei22:nips,li24:ijcai}, generative learning~\cite{du22:bd,kim24:iclr}, or virtual label distributions derived from hypergraph connectivity~\cite{lee24:web}.

\vspace{1mm}
\noindent\textbf{Representation Learning in Hyperbolic Space.}  
Hyperbolic space is a non-Euclidean manifold with negative curvature whose volume grows exponentially with radius, making it well suited for representing hierarchical structures~\cite{liu19:nips,gulcehre19:iclr}.
Since standard Euclidean operations cannot be directly applied to hyperbolic space, existing hyperbolic GRL methods commonly use logarithmic maps to project node representations onto Euclidean tangent spaces, where linear transformations and neighborhood aggregation are performed, and then map the resulting representations back to hyperbolic space using exponential maps~\cite{zhang25:kdd, liu19:nips, liu21:nips}.
Representative methods include HB-GCN~\cite{chami19:nips}, HB-GAT~\cite{gulcehre19:iclr}, and HB-GCL~\cite{zhang25:kdd}, which extend graph convolution, attention, and contrastive learning to hyperbolic space, respectively.
% To avoid confusion with HRL methods, we refer to these methods as HB-GCN, HB-GAT, and HB-GCL, respectively.

More recently, hyperbolic representation learning has been extended from graphs to hypergraphs.
Specifically, THINK~\cite{agarwal22:icdm} learns hyperbolic representations of temporal hypergraphs by performing message passing based on hyperbolic distances between node and hyperedge embeddings.
However, the structural properties of hyperbolic space have received relatively limited attention in HB-HRL, particularly regarding which structural properties can provide effective guidance for hypergraph representation learning.

%% file: s3.tex
\section{Preliminary Analysis}\label{s3}
Hyperbolic space has geometric properties well suited for representing tree-like structures, and hyperbolic representation learning has been primarily studied on graphs.
However, it remains unclear whether hypergraphs exhibit tree-like structures suitable for hyperbolic representation learning and whether simply extending existing hypergraph encoders to hyperbolic space can improve performance over their Euclidean counterparts.
Moreover, unlike graphs with pairwise connectivity, hypergraphs exhibit groupwise connectivity, where multiple nodes interact through a single hyperedge.
This structural difference raises the question of whether structural properties identified in graphs remain effective for guiding hyperbolic representation learning on hypergraphs.
To systematically investigate these issues and identify structural properties suitable for effective HB-HRL, we conduct preliminary experiments guided by the following five preliminary research questions \textbf{(PQs)}.

\begin{itemize}[leftmargin=*]
    \item \textbf{PQ1}: Do hypergraphs exhibit strong tree-like structures suitable for hyperbolic representation learning?
    \item \textbf{PQ2}: Is simply extending existing hypergraph encoders to hyperbolic space sufficient to improve downstream performance?
    \item \textbf{PQ3}: Which node structural properties can provide effective guidance for HB-GRL?
    \item \textbf{PQ4}: Does the same structural guidance remain effective when applied to hypergraphs?
    \item \textbf{PQ5}: Why does the effectiveness of structural guidance differ between graphs and hypergraphs, and what hypergraph-specific structural properties should be considered for HB-HRL?
\end{itemize}

\subsection{PQ1. Suitability of Hyperbolic Space for HRL}\label{s3.1}
\input{Figs/f1}

To assess the suitability of hyperbolic space for HRL, we measure hyperbolicity ($\delta_{hg}$) on four graph datasets used in prior HB-GRL studies~\cite{chami19:nips,zhang25:kdd} and ten hypergraph datasets widely used in HRL~\cite{lee23:aaai,kim24:iclr}, using established graph-specific~\cite{chami19:nips} and hypergraph-specific~\cite{agarwal22:icdm} measures.
Lower hyperbolicity indicates a stronger tree-like structure.
Fig.~\ref{f1} presents the distribution of hyperbolicity across graph and hypergraph datasets, where the \textit{x}-axis represents the graph and hypergraph categories and the \textit{y}-axis represents hyperbolicity ($\delta_{hg}$).
The results show that graph hyperbolicity varies widely from approximately 0 to 11, whereas hypergraphs exhibit relatively low values ranging from 1 to 3.5, despite the two categories having similar median values.
These results indicate that hypergraphs consistently exhibit strong tree-like structures, supporting their geometric suitability for hyperbolic representation learning.
However, whether this geometric suitability translates into improved downstream performance remains unclear.
\vspace{-1mm}
\subsection{PQ2. Effectiveness of Hyperbolic Extension}\label{s3.2}
\vspace{-1mm}
To examine whether hyperbolic space improves HRL performance, we extend four existing HRL models to hyperbolic space using mapping-based operations adopted in HB-GRL, which perform message passing in Euclidean tangent spaces~\cite{chami19:nips}.
Specifically, we select HGNN~\cite{fen19:aaai} and HNHN~\cite{don20:icmlw} as semi-supervised methods based on clique and star expansion, respectively, and TriCL~\cite{lee23:aaai} and HypeBoy~\cite{kim24:iclr} as self-supervised methods based on contrastive and generative learning, respectively.
Table~\ref{t1} reports the node classification accuracy of the original Euclidean HRL models and their hyperbolic counterparts across ten hypergraph datasets, together with the relative performance changes achieved by the hyperbolic variants.
The results show that the effectiveness of hyperbolic extensions varies substantially across models and datasets.
Specifically, HB-HGNN improves HGNN by 18.96\% on Cora-CA and 11.18\% on 20News, while resulting in an 8.20\% decrease on Cora.
Similarly, HB-HNHN achieves an 8.25\% improvement on Pubmed but a 7.24\% degradation on Cora.
Overall, these results demonstrate that simply extending existing HRL models to hyperbolic space does not consistently improve downstream performance, even on hypergraphs with strong tree-like structures.

\input{Tabs/t1}
\input{Tabs/t2}
\input{Figs/f2}
\input{Figs/f3}

\input{Tabs/t3}
\input{Figs/f4}
\vspace{-1mm}
\subsection{PQ3. Structural Guidance for HB-GRL}\label{s3.3}
\vspace{-1mm}
Previous HB-GRL studies have reported consistent performance improvements over Euclidean methods~\cite{chami19:nips,dai21:cvpr,gulcehre19:iclr,zhang25:kdd,liu21:nips}.
To investigate which node structural properties can provide effective guidance for HB-GRL, we analyze node embeddings learned by HB-GCN~\cite{chami19:nips} with respect to nine structural properties: degree, clustering coefficient, and coreness as local-level properties; closeness, betweenness, PageRank, and eigenvector centrality as global-level properties; and label entropy and homophily as community-level properties~\cite{henderson11:kdd,henderson12:kdd,watts98:nature,kitsak10:nphys,platonov23:nips}.
We sort nodes by their embedding radii from the center to the periphery, divide them into ten equal-sized groups, and compute the average value of each structural property within each group.

Fig.~\ref{f2} presents how the nine structural properties vary with node radial positions, where the \textit{x}-axis represents the ten node groups ordered from the center to the periphery and the \textit{y}-axis represents the average value of each structural property within each group.
The results show that only the community-level properties exhibit consistent trends across graph datasets: label entropy decreases while homophily increases from the center toward the periphery.
These trends indicate that nodes closer to the center tend to have more diverse neighborhood labels and lower homophily, characteristics associated with bridge nodes that connect different label groups.
Thus, the results reveal a consistent relationship between node bridging roles and radial positions in hyperbolic space: nodes with stronger bridging roles tend to be positioned closer to the center.

Based on this observation, we quantify such bridging roles using bridge scores and examine whether they can serve as effective structural guidance for HB-GRL.
We incorporate a \textit{bridge-guided loss} into HB-GCN~\cite{chami19:nips}, which encourages nodes with higher bridge scores to have smaller embedding radii while retaining the original node classification objective.
Table~\ref{t2} compares the node classification accuracy of HB-GCN with and without the bridge-guided loss.
The results show that incorporating bridge guidance consistently improves node classification performance across graph datasets.
These results demonstrate that bridge scores can serve as effective structural guidance for organizing node representations along the radial dimension in HB-GRL.
\vspace{-1mm}
\subsection{PQ4. Generalization to Hypergraphs}\label{s3.4}
To examine whether the bridge--radius relationship observed in HB-GRL also holds for HB-HRL, we learn node embeddings using HB-HNHN and repeat the same analysis on hypergraph datasets.
Fig.~\ref{f3} presents the relationship between node radial positions and the two community-level structural properties, label entropy and homophily.
Unlike HB-GRL, HB-HRL exhibits no consistent relationship between radial positions and either label entropy or homophily across hypergraph datasets.

We further examine whether bridge scores can provide effective structural guidance for HB-HRL by applying the same bridge-guided loss used in HB-GRL to HB-HNHN.
Table~\ref{t3} compares the node classification accuracy of HB-HNHN with and without the bridge-guided loss across hypergraph datasets.
The results show that bridge-guided learning does not consistently improve node classification performance, yielding improvements on some datasets but degradation on others.
These results indicate that neither the bridge--radius relationship nor the effectiveness of bridge guidance observed in HB-GRL directly generalizes to HB-HRL.
\subsection{PQ5. Hypergraph-Specific Structural Properties}\label{s3.5}
To investigate why conventional bridge score estimation may be less effective in hypergraphs, we examine if bridge nodes have neighborhood label distributions different from those of their surrounding nodes. Specifically, for each node $v_i$, we quantify how different its neighborhood label distribution is from those of its connected nodes $v_j$ (i.e., neighbors in graphs or hyperedge co-members in hypergraphs) by measuring the difference for each pair $(v_i,v_j)$ and averaging these differences for each $v_i$.
We use Jensen--Shannon Divergence (JSD) to measure these differences, where lower values indicate greater similarity. To examine whether these differences distinguish bridge from non-bridge nodes, we classify a node as a bridge node if it connects to at least one node with a different label, and as a non-bridge node otherwise.
\input{Figs/f5}

Fig.~\ref{f4} presents the JSD distributions of bridge and non-bridge nodes across graph and hypergraph datasets, where the \textit{x}-axis represents the datasets and the \textit{y}-axis represents JSD.
In graph datasets, bridge nodes consistently exhibit higher JSD values than non-bridge nodes, showing a clear distinction between the two groups.
In hypergraph datasets, bridge nodes generally exhibit higher JSD values, but their distributions vary substantially and frequently extend toward near-zero JSD values.
This indicates that many bridge nodes can still have neighborhood label distributions highly similar to those of their hyperedge co-members, making their bridging roles difficult to distinguish using conventional node-level bridge score estimation.
Moreover, conventional bridge scores aggregate all neighboring nodes without accounting for the distinct bridging roles of individual hyperedges.
These findings suggest that effective bridge score estimation for hypergraphs should explicitly account for how individual hyperedges connect different external node groups.

%% file: Figs/f1.tex
\begin{figure}[t]
\centering
\begin{tikzpicture}
\begin{axis}[
    width=7.5cm,
    height=5.0cm,
    ymin=-0.5, ymax=11.5,
    xmin=0.5, xmax=2.5,
    ylabel={Hyperbolicity($\delta_{hg}$)},
    title={Graph vs.\ Hypergraph},
    xtick={1,2},
    xticklabels={Graph, Hypergraph},
    xticklabel style={font=\footnotesize},
    yticklabel style={font=\footnotesize},
    label style={font=\footnotesize},
    title style={font=\footnotesize},
    ymajorgrids=true,
    grid style={dashed, gray!25},
    boxplot/draw direction=y,
    boxplot/box extend=0.35,
    axis x line*=bottom,
    axis y line*=left,
]

% Graph
\addplot+[
    boxplot prepared={
        draw position=1,
        lower whisker=0,
        lower quartile=0.75,
        median=2.25,
        upper quartile=5.35,
        upper whisker=11
    },
    boxplot/every box/.style={
        fill=red!35,
        fill opacity=0.6,
        draw=none
    },
    boxplot/every median/.style={
        draw=black,
        line width=1pt
    },
    boxplot/every whisker/.style={
        draw=black,
        line width=0.8pt
    },
] coordinates {};

% Hypergraph
\addplot+[
    boxplot prepared={
        draw position=2,
        lower whisker=1,
        lower quartile=2,
        median=2.25,
        upper quartile=2.75,
        upper whisker=3.5
    },
    boxplot/every box/.style={
        fill=cyan!45,
        fill opacity=0.7,
        draw=none
    },
    boxplot/every median/.style={
        draw=black,
        line width=1pt
    },
    boxplot/every whisker/.style={
        draw=black,
        line width=0.8pt
    },
] coordinates {};

\end{axis}
\end{tikzpicture}

\vspace{-0.2cm}
\caption{Hyperbolicity($\delta_{hg}$) of graphs and hypergraphs.}
\label{f1}
\vspace{-5mm}
\end{figure}
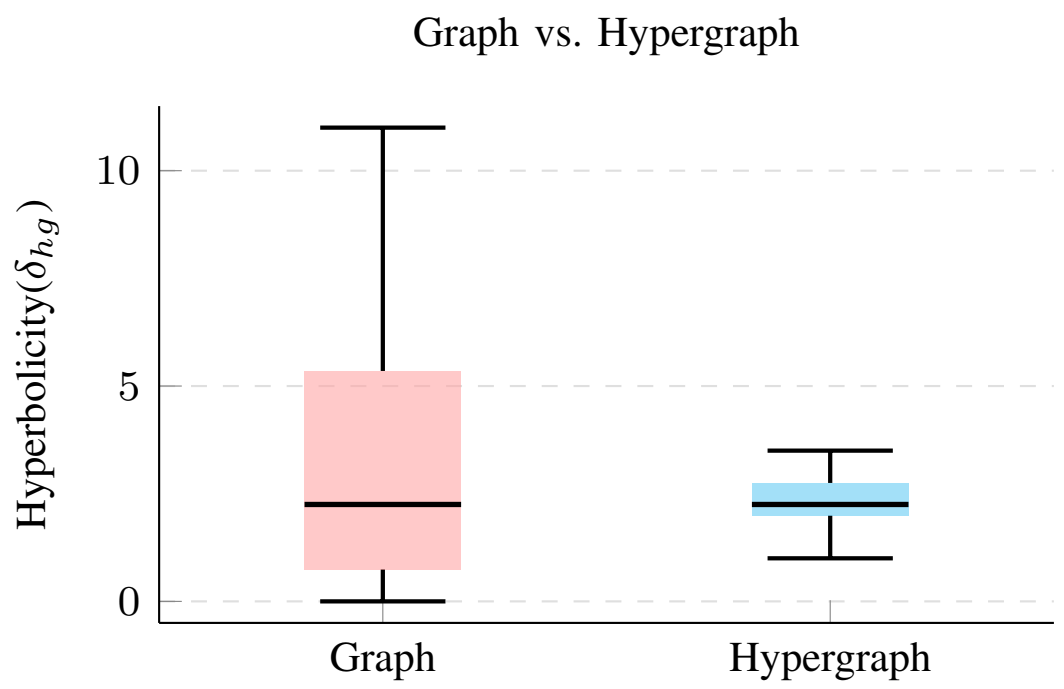

%% file: Tabs/t1.tex
\begin{table*}[t]
\centering
\caption{Node classification accuracy of Euclidean and hyperbolic
HRL methods}
\label{t1}
\footnotesize
\setlength{\tabcolsep}{3pt}
\renewcommand{\arraystretch}{1.2}

\resizebox{\textwidth}{!}{%
\begin{tabular}{lcccccccccc}
\toprule[1.2pt]
\textbf{Method}
& \textbf{Citeseer}
& \textbf{Cora}
& \textbf{DBLP-A}
& \textbf{Cora-CA}
& \textbf{IMDB}
& \textbf{Pubmed}
& \textbf{DBLP-P}
& \textbf{Aminer}
& \textbf{MN-40}
& \textbf{20News} \\
\midrule[1.0pt]

% ================= HGNN =================
HGNN
& 38.1 $\pm$ 10.7
& 45.1 $\pm$ 9.1
& 65.0 $\pm$ 5.7
& 44.3 $\pm$ 8.0
& 41.5 $\pm$ 2.8
& 70.8 $\pm$ 3.8
& 84.0 $\pm$ 0.9
& 29.7 $\pm$ 2.0
& 89.5 $\pm$ 1.3
& 67.1 $\pm$ 2.0 \\

HB-HGNN
& 36.8 $\pm$ 7.1
& 41.4 $\pm$ 5.7
& 64.3 $\pm$ 3.5
& 52.7 $\pm$ 6.7
& 40.1 $\pm$ 1.8
& 71.9 $\pm$ 4.0
& 85.7 $\pm$ 0.8
& 30.3 $\pm$ 1.1
& 86.0 $\pm$ 2.1
& 74.6 $\pm$ 0.8 \\

\midrule[0.4pt]
\textbf{Improvement}
& \textcolor{red}{\textbf{-3.41\%}}
& \textcolor{red}{\textbf{-8.20\%}}
& \textcolor{red}{\textbf{-1.08\%}}
& \textcolor{blue}{\textbf{+18.96\%}}
& \textcolor{red}{\textbf{-3.37\%}}
& \textcolor{blue}{\textbf{+1.55\%}}
& \textcolor{blue}{\textbf{+2.02\%}}
& \textcolor{blue}{\textbf{+2.02\%}}
& \textcolor{red}{\textbf{-3.91\%}}
& \textcolor{blue}{\textbf{+11.18\%}} \\

\midrule[1.0pt]

% ================= HNHN =================
HNHN
& 44.2 $\pm$ 7.9
& 52.5 $\pm$ 7.3
& 69.0 $\pm$ 4.5
& 53.1 $\pm$ 7.4
& 42.5 $\pm$ 3.6
& 69.1 $\pm$ 3.4
& 86.4 $\pm$ 0.8
& 32.1 $\pm$ 2.7
& 89.2 $\pm$ 1.4
& 74.4 $\pm$ 1.1 \\

HB-HNHN
& 42.1 $\pm$ 7.7
& 48.7 $\pm$ 10.8
& 65.2 $\pm$ 5.2
& 53.6 $\pm$ 6.2
& 41.6 $\pm$ 3.9
& 74.8 $\pm$ 4.1
& 86.3 $\pm$ 0.7
& 30.9 $\pm$ 1.5
& 90.1 $\pm$ 0.7
& 76.7 $\pm$ 0.8 \\

\midrule[0.4pt]
\textbf{Improvement}
& \textcolor{red}{\textbf{-4.75\%}}
& \textcolor{red}{\textbf{-7.24\%}}
& \textcolor{red}{\textbf{-5.51\%}}
& \textcolor{blue}{\textbf{+0.94\%}}
& \textcolor{red}{\textbf{-2.12\%}}
& \textcolor{blue}{\textbf{+8.25\%}}
& \textcolor{red}{\textbf{-0.12\%}}
& \textcolor{red}{\textbf{-3.74\%}}
& \textcolor{blue}{\textbf{+1.01\%}}
& \textcolor{blue}{\textbf{+3.09\%}} \\

\midrule[1.0pt]

% ================= TriCL =================
TriCL
& 53.0 $\pm$ 9.9
& 61.5 $\pm$ 5.3
& 80.3 $\pm$ 3.5
& 63.4 $\pm$ 3.9
& 47.5 $\pm$ 4.6
& 74.0 $\pm$ 5.2
& 87.6 $\pm$ 0.5
& 34.6 $\pm$ 2.7
& 93.0 $\pm$ 0.8
& 73.8 $\pm$ 1.8 \\

HB-TriCL
& 55.8 $\pm$ 8.7
& 59.7 $\pm$ 6.1
& 78.9 $\pm$ 3.9
& 64.2 $\pm$ 4.3
& 46.8 $\pm$ 4.2
& 76.1 $\pm$ 4.7
& O.O.M
& O.O.M
& 92.5 $\pm$ 1.0
& 75.4 $\pm$ 1.5 \\

\midrule[0.4pt]
\textbf{Improvement}
& \textcolor{blue}{\textbf{+5.28\%}}
& \textcolor{red}{\textbf{-2.93\%}}
& \textcolor{red}{\textbf{-1.74\%}}
& \textcolor{blue}{\textbf{+1.26\%}}
& \textcolor{red}{\textbf{-1.47\%}}
& \textcolor{blue}{\textbf{+2.84\%}}
& --
& --
& \textcolor{red}{\textbf{-0.54\%}}
& \textcolor{blue}{\textbf{+2.17\%}} \\

\midrule[1.0pt]

% ================= Hypeboy =================
Hypeboy
& 57.7 $\pm$ 10.8
& 61.9 $\pm$ 8.5
& 81.2 $\pm$ 2.7
& 67.0 $\pm$ 3.7
& 48.3 $\pm$ 5.0
& 73.7 $\pm$ 4.1
& 87.9 $\pm$ 0.5
& 34.6 $\pm$ 2.8
& 89.2 $\pm$ 2.2
& 75.7 $\pm$ 2.1 \\

HB-Hypeboy
& 58.0 $\pm$ 8.1
& 63.1 $\pm$ 1.5
& 79.0 $\pm$ 3.5
& 64.0 $\pm$ 1.0
& 50.0 $\pm$ 1.7
& 71.6 $\pm$ 2.0
& O.O.M
& 34.9 $\pm$ 1.8
& 93.3 $\pm$ 1.2
& 73.8 $\pm$ 2.6 \\

\midrule[0.4pt]
\textbf{Improvement}
& \textcolor{blue}{\textbf{+0.52\%}}
& \textcolor{blue}{\textbf{+1.94\%}}
& \textcolor{red}{\textbf{-2.71\%}}
& \textcolor{red}{\textbf{-4.48\%}}
& \textcolor{blue}{\textbf{+3.52\%}}
& \textcolor{red}{\textbf{-2.85\%}}
& --
& \textcolor{blue}{\textbf{+0.87\%}}
& \textcolor{blue}{\textbf{+4.60\%}}
& \textcolor{red}{\textbf{-2.51\%}} \\

\bottomrule[1.2pt]
\end{tabular}%
}
\end{table*}

%% file: Tabs/t2.tex
\begin{table}[t]
\centering
\caption{Node classification accuracy of HB-GCN with and without bridge-guided learning}
\label{t2}
\scriptsize
\setlength{\tabcolsep}{3.5pt}
\renewcommand{\arraystretch}{1.2}

\resizebox{0.9\columnwidth}{!}{%
\begin{tabular}{lcccc}
\toprule
\textbf{Method}
& \textbf{Cora}
& \textbf{Airport}
& \textbf{Pubmed}
& \textbf{Disease} \\
\midrule

HB-GCN
& 0.787
& 0.826
& 0.738
& 0.763 \\

HB-GCN + Bridge
& \textbf{0.802}
& \textbf{0.870}
& \textbf{0.762}
& \textbf{0.799} \\

\midrule
\textbf{Improvement}
& \textcolor{blue}{\textbf{+1.91\%}}
& \textcolor{blue}{\textbf{+5.33\%}}
& \textcolor{blue}{\textbf{+3.25\%}}
& \textcolor{blue}{\textbf{+4.72\%}} \\

\bottomrule
\end{tabular}%
}
\vspace{-8mm}
\end{table}

%% file: Figs/f2.tex
% Required packages (add to your existing preamble):
% \usepackage{graphicx} % for \resizebox
% \usepackage{pgfplots}
% \usepgfplotslibrary{groupplots}
% \pgfplotsset{compat=1.18}
%
% This figure uses only \addplot coordinates and groupplot.
% No pgfplotstable/statistics or external data files required.
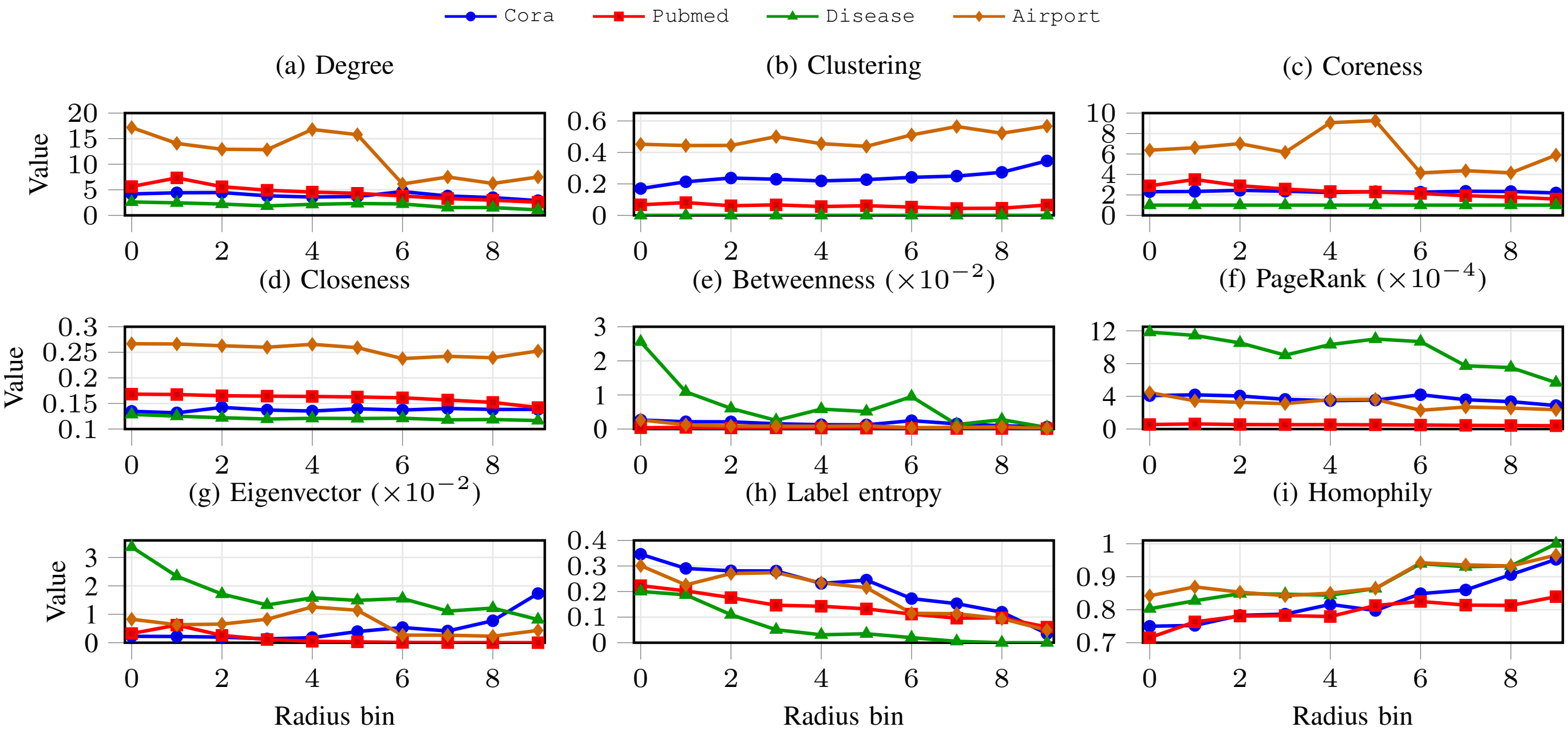
\begin{figure*}[t]
    \centering
    \scriptsize

    % Shared legend; compile twice if it is missing on the first pass.
    \pgfplotslegendfromname{radialNineLegend}
    \vspace{1mm}

    \resizebox{\textwidth}{!}{%
    \begin{tikzpicture}
    \begin{groupplot}[
        group style={
            group size=3 by 3,
            horizontal sep=0.80cm,
            vertical sep=1.00cm
        },
        width=0.295\textwidth,
        height=2.5cm,
        xmin=-0.15, xmax=9.15,
        xtick={0,2,4,6,8},
        grid=major,
        grid style={gray!18},
        tick align=outside,
        tick pos=left,
        scaled y ticks=false,
        axis line style={line width=0.65pt},
        label style={font=\scriptsize},
        tick label style={font=\scriptsize, /pgf/number format/fixed, /pgf/number format/precision=2},
        title style={font=\scriptsize},
        every axis plot/.append style={
            line width=0.90pt,
            mark size=1.15pt
        },
        legend columns=4,
        legend style={
            draw=none,
            font=\scriptsize,
            /tikz/every even column/.append style={column sep=0.35cm}
        },
        legend cell align=left
    ]

    % ---------------- (a) Degree ----------------
    \nextgroupplot[
        title={(a) Degree},
        ymin=0, ymax=20, ytick={0,5,10,15,20},
        ylabel={Value},
        legend to name=radialNineLegend
    ]
    % Cora
    \addplot+[blue, mark=*] coordinates {
        (0,4.19557196) (1,4.41328413) (2,4.46125461) (3,3.82656827) (4,3.58302583)
        (5,3.69372694) (6,4.65313653) (7,3.7896679) (8,3.47407407) (9,2.88518519)
    };
    \addlegendentry{\texttt{Cora}}
    % Pubmed
    \addplot+[red, mark=square*] coordinates {
        (0,5.63336714) (1,7.32758621) (2,5.60750507) (3,4.9137931) (4,4.55172414)
        (5,4.31490872) (6,3.78245436) (7,3.29223744) (8,2.96245561) (9,2.57179097)
    };
    \addlegendentry{\texttt{Pubmed}}
    % Disease
    \addplot+[green!60!black, mark=triangle*] coordinates {
        (0,2.61904762) (1,2.44761905) (2,2.20952381) (3,1.84761905) (4,2.16346154)
        (5,2.31730769) (6,2.23076923) (7,1.53846154) (8,1.51923077) (9,1.07692308)
    };
    \addlegendentry{\texttt{Disease}}
    % Airport
    \addplot+[orange!80!black, mark=diamond*] coordinates {
        (0,17.17868339) (1,14.06583072) (2,12.92476489) (3,12.84952978) (4,16.79874214)
        (5,15.77987421) (6,6.13836478) (7,7.50628931) (8,6.24213836) (9,7.50628931)
    };
    \addlegendentry{\texttt{Airport}}

    % ---------------- (b) Clustering ----------------
    \nextgroupplot[
        title={(b) Clustering},
        ymin=0, ymax=0.65, ytick={0,0.2,0.4,0.6}
    ]
    % Cora
    \addplot+[blue, mark=*] coordinates {
        (0,0.1702778) (1,0.21339935) (2,0.23725268) (3,0.22950693) (4,0.21896834)
        (5,0.22691575) (6,0.24163798) (7,0.2496127) (8,0.27364766) (9,0.34602422)
    };
    % Pubmed
    \addplot+[red, mark=square*] coordinates {
        (0,0.06728834) (1,0.08113837) (2,0.06129151) (3,0.06678926) (4,0.05614833)
        (5,0.06121176) (6,0.0527427) (7,0.04405089) (8,0.04539983) (9,0.06567822)
    };
    % Disease
    \addplot+[green!60!black, mark=triangle*] coordinates {
        (0,0.0) (1,0.0) (2,0.0) (3,0.0) (4,0.0)
        (5,0.0) (6,0.0) (7,0.0) (8,0.0) (9,0.0)
    };
    % Airport
    \addplot+[orange!80!black, mark=diamond*] coordinates {
        (0,0.45192286) (1,0.44380836) (2,0.4443419) (3,0.50026254) (4,0.45567491)
        (5,0.43867907) (6,0.51117944) (7,0.56444661) (8,0.52182093) (9,0.56677676)
    };

    % ---------------- (c) Coreness ----------------
    \nextgroupplot[
        title={(c) Coreness},
        ymin=0, ymax=10, ytick={0,2,4,6,8,10}
    ]
    % Cora
    \addplot+[blue, mark=*] coordinates {
        (0,2.31365314) (1,2.33579336) (2,2.4501845) (3,2.35793358) (4,2.2398524)
        (5,2.31734317) (6,2.27306273) (7,2.35793358) (8,2.33333333) (9,2.19259259)
    };
    % Pubmed
    \addplot+[red, mark=square*] coordinates {
        (0,2.89249493) (1,3.50709939) (2,2.89756592) (3,2.5796146) (4,2.3331643)
        (5,2.28549696) (6,2.13843813) (7,1.92541857) (8,1.78792491) (9,1.5971588)
    };
    % Disease
    \addplot+[green!60!black, mark=triangle*] coordinates {
        (0,1.0) (1,1.0) (2,1.0) (3,1.0) (4,1.0)
        (5,1.0) (6,1.0) (7,1.0) (8,1.0) (9,1.0)
    };
    % Airport
    \addplot+[orange!80!black, mark=diamond*] coordinates {
        (0,6.36990596) (1,6.61128527) (2,7.00940439) (3,6.1630094) (4,9.05974843)
        (5,9.25786164) (6,4.1509434) (7,4.36477987) (8,4.1572327) (9,5.89308176)
    };

    % ---------------- (d) Closeness ----------------
    \nextgroupplot[
        title={(d) Closeness},
        ymin=0.1, ymax=0.3, ytick={0.10,0.15,0.20,0.25,0.30},
        ylabel={Value}
    ]
    % Cora
    \addplot+[blue, mark=*] coordinates {
        (0,0.13467082) (1,0.13169787) (2,0.14234644) (3,0.13715689) (4,0.13516738)
        (5,0.13959811) (6,0.13713839) (7,0.14022579) (8,0.13834841) (9,0.13841688)
    };
    % Pubmed
    \addplot+[red, mark=square*] coordinates {
        (0,0.16819663) (1,0.16748261) (2,0.16507513) (3,0.16434511) (4,0.16372336)
        (5,0.16269971) (6,0.16106454) (7,0.15670608) (8,0.152106) (9,0.14201084)
    };
    % Disease
    \addplot+[green!60!black, mark=triangle*] coordinates {
        (0,0.12864892) (1,0.12517264) (2,0.12211845) (3,0.11954669) (4,0.12095129)
        (5,0.12056749) (6,0.12113138) (7,0.11810428) (8,0.11864719) (9,0.11660624)
    };
    % Airport
    \addplot+[orange!80!black, mark=diamond*] coordinates {
        (0,0.26673118) (1,0.26626969) (2,0.2628048) (3,0.25996464) (4,0.26562495)
        (5,0.25908735) (6,0.23774855) (7,0.24223924) (8,0.23934197) (9,0.25264969)
    };

    % ---------------- (e) Betweenness ($\times 10^{-2}$) ----------------
    \nextgroupplot[
        title={(e) Betweenness ($\times 10^{-2}$)},
        ymin=0, ymax=0.03, ytick={0,0.01,0.02,0.03}, yticklabels={0,1,2,3}
    ]
    % Cora
    \addplot+[blue, mark=*] coordinates {
        (0,0.00265704) (1,0.002143) (2,0.00211498) (3,0.00155405) (4,0.00127376)
        (5,0.00124059) (6,0.00245415) (7,0.00152403) (8,0.00097948) (9,0.00058009)
    };
    % Pubmed
    \addplot+[red, mark=square*] coordinates {
        (0,0.00039307) (1,0.00045906) (2,0.00032882) (3,0.00034126) (4,0.00029953)
        (5,0.00027208) (6,0.00019245) (7,0.00016136) (8,0.00015011) (9,0.00010907)
    };
    % Disease
    \addplot+[green!60!black, mark=triangle*] coordinates {
        (0,0.0255787) (1,0.01090051) (2,0.00605172) (3,0.0025418) (4,0.0058499)
        (5,0.00514174) (6,0.00951343) (7,0.00132355) (8,0.00275269) (9,0.00045824)
    };
    % Airport
    \addplot+[orange!80!black, mark=diamond*] coordinates {
        (0,0.00265838) (1,0.001258) (2,0.00090177) (3,0.00084047) (4,0.00083148)
        (5,0.00093929) (6,0.0003937) (7,0.00050256) (8,0.00065817) (9,0.00016296)
    };

    % ---------------- (f) PageRank ($\times 10^{-4}$) ----------------
    \nextgroupplot[
        title={(f) PageRank ($\times 10^{-4}$)},
        ymin=0, ymax=0.00125, ytick={0,0.0004,0.0008,0.0012}, yticklabels={0,4,8,12}
    ]
    % Cora
    \addplot+[blue, mark=*] coordinates {
        (0,0.00040857) (1,0.00041797) (2,0.00040407) (3,0.00036268) (4,0.00034667)
        (5,0.00035369) (6,0.00041904) (7,0.00035789) (8,0.00033474) (9,0.000287)
    };
    % Pubmed
    \addplot+[red, mark=square*] coordinates {
        (0,5.51988870e-05) (1,6.23711375e-05) (2,5.51289308e-05) (3,5.32998662e-05) (4,5.32043395e-05)
        (5,5.15253080e-05) (6,4.84252445e-05) (7,4.52510926e-05) (8,4.28231185e-05) (9,3.99363810e-05)
    };
    % Disease
    \addplot+[green!60!black, mark=triangle*] coordinates {
        (0,0.0011838) (1,0.00114405) (2,0.00105212) (3,0.00090327) (4,0.00103247)
        (5,0.00110003) (6,0.00106803) (7,0.00077182) (8,0.00075133) (9,0.00056728)
    };
    % Airport
    \addplot+[orange!80!black, mark=diamond*] coordinates {
        (0,0.00044517) (1,0.00034552) (2,0.00032614) (3,0.00031085) (4,0.00035884)
        (5,0.0003628) (6,0.00022882) (7,0.00026916) (8,0.00025595) (9,0.00023693)
    };

    % ---------------- (g) Eigenvector ($\times 10^{-2}$) ----------------
    \nextgroupplot[
        title={(g) Eigenvector ($\times 10^{-2}$)},
        ymin=0, ymax=0.036, ytick={0,0.01,0.02,0.03}, yticklabels={0,1,2,3},
        ylabel={Value},
        xlabel={Radius bin}
    ]
    % Cora
    \addplot+[blue, mark=*] coordinates {
        (0,0.0022417) (1,0.00218969) (2,0.00195594) (3,0.00133903) (4,0.00178277)
        (5,0.00392491) (6,0.00534246) (7,0.00411823) (8,0.00772016) (9,0.01730759)
    };
    % Pubmed
    \addplot+[red, mark=square*] coordinates {
        (0,3.24017740e-03) (1,6.19216414e-03) (2,2.58340723e-03) (3,1.11508041e-03) (4,4.94618791e-04)
        (5,3.01020242e-04) (6,9.82242000e-05) (7,4.00685122e-05) (8,2.60490383e-05) (9,8.11859280e-06)
    };
    % Disease
    \addplot+[green!60!black, mark=triangle*] coordinates {
        (0,0.03367778) (1,0.02337997) (2,0.01711919) (3,0.01328176) (4,0.01579924)
        (5,0.014893) (6,0.01554538) (7,0.01111909) (8,0.01218447) (9,0.00813633)
    };
    % Airport
    \addplot+[orange!80!black, mark=diamond*] coordinates {
        (0,0.00825433) (1,0.00633717) (2,0.0065387) (3,0.00825855) (4,0.01258065)
        (5,0.01141616) (6,0.00266864) (7,0.00264036) (8,0.00228206) (9,0.00432422)
    };

    % ---------------- (h) Label entropy ----------------
    \nextgroupplot[
        title={(h) Label entropy},
        ymin=0, ymax=0.40, ytick={0,0.10,0.20,0.30,0.40},
        xlabel={Radius bin}
    ]
    % Cora
    \addplot+[blue, mark=*] coordinates {
        (0,0.34597493) (1,0.29037363) (2,0.28100962) (3,0.28039572) (4,0.23164593)
        (5,0.2450582) (6,0.17248567) (7,0.15265316) (8,0.11937437) (9,0.03543395)
    };
    % Pubmed
    \addplot+[red, mark=square*] coordinates {
        (0,0.22314529) (1,0.20260497) (2,0.17616935) (3,0.14640018) (4,0.14254386)
        (5,0.13229813) (6,0.11186713) (7,0.09597757) (8,0.09713086) (9,0.06091819)
    };
    % Disease
    \addplot+[green!60!black, mark=triangle*] coordinates {
        (0,0.20052045) (1,0.18708068) (2,0.10965404) (3,0.05059545) (4,0.03098618)
        (5,0.03498134) (6,0.0198243) (7,0.00540707) (8,0.0) (9,0.0)
    };
    % Airport
    \addplot+[orange!80!black, mark=diamond*] coordinates {
        (0,0.3014055) (1,0.22598826) (2,0.27044801) (3,0.2739108) (4,0.23400368)
        (5,0.21531085) (6,0.11497447) (7,0.11334217) (8,0.09410216) (9,0.04911299)
    };

    % ---------------- (i) Homophily ----------------
    \nextgroupplot[
        title={(i) Homophily},
        ymin=0.70, ymax=1.01, ytick={0.70,0.80,0.90,1.00},
        xlabel={Radius bin}
    ]
    % Cora
    \addplot+[blue, mark=*] coordinates {
        (0,0.74982056) (1,0.75228551) (2,0.78263536) (3,0.78665012) (4,0.81604778)
        (5,0.79755386) (6,0.84862339) (7,0.85985377) (8,0.90634664) (9,0.95253086)
    };
    % Pubmed
    \addplot+[red, mark=square*] coordinates {
        (0,0.71511479) (1,0.76313921) (2,0.78068681) (3,0.78210769) (4,0.7792539)
        (5,0.81271633) (6,0.82489776) (7,0.81396012) (8,0.81281755) (9,0.83947412)
    };
    % Disease
    \addplot+[green!60!black, mark=triangle*] coordinates {
        (0,0.8029932) (1,0.82714286) (2,0.84861678) (3,0.84698413) (4,0.84404762)
        (5,0.86414835) (6,0.93944597) (7,0.93028846) (8,0.93269231) (9,1.0)
    };
    % Airport
    \addplot+[orange!80!black, mark=diamond*] coordinates {
        (0,0.84219819) (1,0.86918286) (2,0.85309367) (3,0.842507) (4,0.85022197)
        (5,0.8645633) (6,0.94215113) (7,0.93566429) (8,0.93243204) (9,0.9653755)
    };

    \end{groupplot}
    \end{tikzpicture}%
    }

    \vspace{-2mm}
    \caption{Graph structural properties across
    hyperbolic radius bins.}
    \label{f2}
    \vspace{-3mm}
\end{figure*}

%% file: Figs/f3.tex
\begin{figure}[t]
    \centering
    \scriptsize

    \begin{tikzpicture}
    \begin{groupplot}[
        group style={
            group size=2 by 1,
            horizontal sep=0.75cm,
            vertical sep=1.0cm
        },
        width=0.27\textwidth,
        height=2.4cm,
        xmin=0, xmax=9,
        xtick={0,1,2,3,4,5,6,7,8,9},
        grid=both,
        grid style={gray!20},
        tick align=outside,
        tick pos=left,
        label style={font=\scriptsize},
        tick label style={font=\scriptsize},
        every axis plot/.append style={
            line width=0.85pt,
            mark size=0.45pt
        },
        legend columns=3,
        legend style={
            at={(1.08,1.45)},
            anchor=south,
            draw=none,
            font=\scriptsize,
            /tikz/every even column/.append style={
                column sep=0.25cm
            }
        },
        legend cell align=left,
        clip=false
    ]

    % =============== Label entropy ===============
    \nextgroupplot[
        ylabel={Value},
        ymin=0, ymax=0.85
    ]

    % Citeseer
    \addplot+[blue, mark=*] coordinates {
        (0,0.3016645) (1,0.32584016)
        (2,0.28010039) (3,0.10776777)
        (4,0.099319186) (5,0.13661451)
        (6,0.11507831) (7,0.12949497)
        (8,0.1516372) (9,0.13913892)
    };
    \addlegendentry{\texttt{Citeseer}}

    % Cora
    \addplot+[red, mark=square*] coordinates {
        (0,0.14468529) (1,0.33619063)
        (2,0.35130552) (3,0.27139366)
        (4,0.17603556) (5,0.10001172)
        (6,0.081875566) (7,0.091430697)
        (8,0.076436242) (9,0.079421649)
    };
    \addlegendentry{\texttt{Cora}}

    % Cora-CA
    \addplot+[green!60!black, mark=triangle*] coordinates {
        (0,0.3796308) (1,0.2987607)
        (2,0.15597838) (3,0.17975379)
        (4,0.17801057) (5,0.20216383)
        (6,0.18105117) (7,0.14376314)
        (8,0.14328988) (9,0.15773546)
    };
    \addlegendentry{\texttt{Cora-CA}}

    % DBLP-A
    \addplot+[orange!80!black, mark=diamond*] coordinates {
        (0,0.05952614) (1,0.34234393)
        (2,0.25442798) (3,0.15365915)
        (4,0.06455776) (5,0.031609475)
        (6,0.033736601) (7,0.042080828)
        (8,0.039129917) (9,0.035714286)
    };
    \addlegendentry{\texttt{DBLP-A}}

    % IMDB
    \addplot+[purple, mark=pentagon*] coordinates {
        (0,0.63085782) (1,0.76100991)
        (2,0.78853699) (3,0.79328468)
        (4,0.76978576) (5,0.77437905)
        (6,0.76150415) (7,0.74755203)
        (8,0.69211229) (9,0.55473327)
    };
    \addlegendentry{\texttt{IMDB}}

    % Pubmed
    \addplot+[cyan!70!black, mark=asterisk] coordinates {
        (0,0.66292612)
        (1,0.59002569) (2,0.50306139)
        (3,0.44366308) (4,0.38892102)
        (5,0.33528648) (6,0.29185502)
        (7,0.30278952) (8,0.31200186)
        (9,0.32037333)
    };
    \addlegendentry{\texttt{Pubmed}}

    \node[font=\scriptsize, anchor=north]
    at (rel axis cs:0.5,-0.5)
    {(a) Label entropy};

    % ================= Homophily =================
    \nextgroupplot[
        ylabel={},
        ymin=0.4, ymax=1.02
    ]

    % Citeseer
    \addplot+[blue, mark=*] coordinates {
        (0,0.55513908) (1,0.63892872)
        (2,0.667265) (3,0.73909971)
        (4,0.76030275) (5,0.65138603)
        (6,0.71333814) (7,0.71940367)
        (8,0.64122518) (9,0.63599068)
    };

    % Cora
    \addplot+[red, mark=square*] coordinates {
        (0,0.71111087) (1,0.67475753)
        (2,0.62521049) (3,0.7120455)
        (4,0.81092566) (5,0.85076817)
        (6,0.8025641) (7,0.76981352)
        (8,0.79079254) (9,0.7995338)
    };

    % Cora-CA
    \addplot+[green!60!black, mark=triangle*] coordinates {
        (0,0.60744456) (1,0.66620724)
        (2,0.75986323) (3,0.75541729)
        (4,0.76701641) (5,0.71938898)
        (6,0.74079139) (7,0.80689011)
        (8,0.79139102) (9,0.78405137)
    };

    % DBLP-A
    \addplot+[orange!80!black, mark=diamond*] coordinates {
        (0,0.88525669) (1,0.70777708)
        (2,0.77432158) (3,0.85706771)
        (4,0.80521872) (5,0.88938224)
        (6,0.84427284) (7,0.86570142)
        (8,0.88146718) (9,0.84298584)
    };

    % IMDB
    \addplot+[purple, mark=pentagon*] coordinates {
        (0,0.45549656) (1,0.48432772)
        (2,0.4396239) (3,0.46055032)
        (4,0.46077154) (5,0.47604104)
        (6,0.46609322) (7,0.44756726)
        (8,0.48243232) (9,0.49813307)
    };

    % Pubmed
    \addplot+[cyan!70!black, mark=asterisk] coordinates {
        (0,0.52300623)
        (1,0.60149372) (2,0.68103007)
        (3,0.71977245) (4,0.77881087)
        (5,0.80912622) (6,0.82496988)
        (7,0.81605798) (8,0.81422125)
        (9,0.78124611)
    };

    \node[font=\scriptsize, anchor=north]
    at (rel axis cs:0.5,-0.5)
    {(b) Homophily};

    \end{groupplot}
    \end{tikzpicture}

    \vspace{-2mm}
    \caption{
        Hypergraph community properties across
        hyperbolic radius bins.
    }
    \label{f3}
    \vspace{-8mm}
\end{figure}
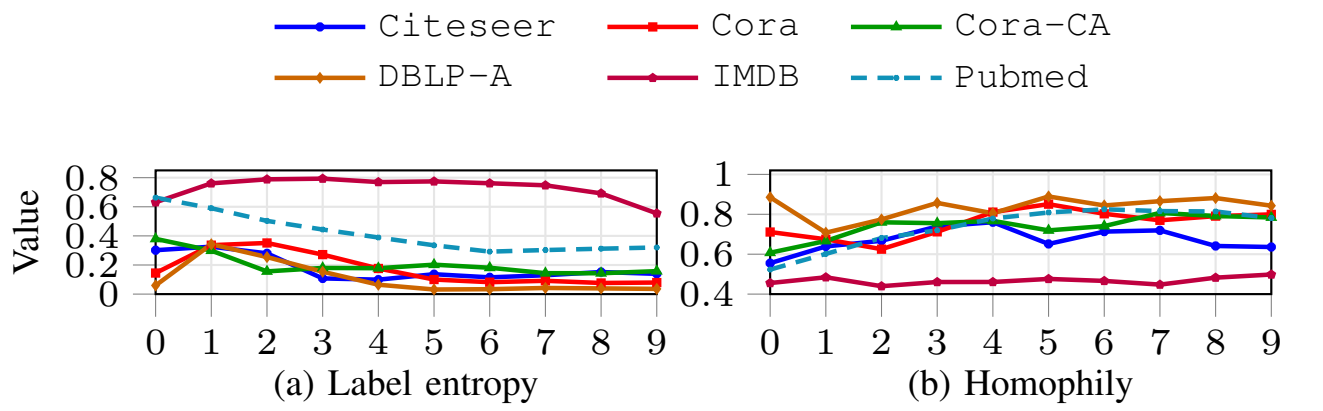

%% file: Tabs/t3.tex
\begin{table*}[t]
\centering
\caption{Node classification accuracy of HB-HNHN with and without bridge-guided learning}
\label{t3}
\footnotesize
\setlength{\tabcolsep}{3pt}
\renewcommand{\arraystretch}{1.25}

\resizebox{\textwidth}{!}{%
\begin{tabular}{lcccccccccc}
\toprule
\textbf{Method}
& \textbf{Citeseer}
& \textbf{Cora}
& \textbf{DBLP-P}
& \textbf{Cora-CA}
& \textbf{IMDB}
& \textbf{Pubmed}
& \textbf{DBLP-A}
& \textbf{Aminer}
& \textbf{MN-40}
& \textbf{20News} \\
\midrule

HB-HNHN
& \textbf{42.1} $\pm$ 7.7
& 48.7 $\pm$ 10.8
& 65.2 $\pm$ 5.2
& 53.6 $\pm$ 6.2
& \textbf{41.6} $\pm$ 3.9
& \textbf{74.8} $\pm$ 4.1
& \textbf{86.3} $\pm$ 0.7
& 30.9 $\pm$ 1.5
& \textbf{90.1} $\pm$ 0.7
& 76.7 $\pm$ 0.8 \\

HB-HNHN + Bridge
& 41.5 $\pm$ 8.1
& \textbf{49.3} $\pm$ 8.8
& \textbf{67.0} $\pm$ 4.1
& \textbf{53.7} $\pm$ 8.6
& 41.5 $\pm$ 4.0
& 74.5 $\pm$ 3.9
& 86.2 $\pm$ 1.0
& \textbf{31.3} $\pm$ 1.6
& 90.0 $\pm$ 0.7
& \textbf{76.8} $\pm$ 1.3 \\

\midrule
\textbf{Improvement}
& \textcolor{red}{\textbf{-1.43\%}}
& \textcolor{blue}{\textbf{+1.23\%}}
& \textcolor{blue}{\textbf{+2.76\%}}
& \textcolor{blue}{\textbf{+0.19\%}}
& \textcolor{red}{\textbf{-0.24\%}}
& \textcolor{red}{\textbf{-0.40\%}}
& \textcolor{red}{\textbf{-0.12\%}}
& \textcolor{blue}{\textbf{+1.29\%}}
& \textcolor{red}{\textbf{-0.11\%}}
& \textcolor{blue}{\textbf{+0.13\%}} \\

\bottomrule
\end{tabular}%
}
\vspace{-2mm}
\end{table*}

%% file: Figs/f4.tex
\begin{figure*}[t]
\centering
\scriptsize

% ================= Colors =================
\definecolor{gNon}{RGB}{166,206,227} % Graph non-bridge
\definecolor{gBri}{RGB}{244,142,81}  % Graph bridge
\definecolor{hHom}{RGB}{145,200,164} % Hypergraph non-bridge
\definecolor{hMix}{RGB}{183,160,217} % Hypergraph bridge

% ================= Boxplot macro =================
% #1: position
% #2: minimum
% #3: Q1
% #4: median
% #5: Q3
% #6: maximum
% #7: fill color

\newcommand{\mybox}[7]{%
    \addplot[
        boxplot prepared={
            draw position=#1,
            lower whisker=#2,
            lower quartile=#3,
            median=#4,
            upper quartile=#5,
            upper whisker=#6
        },
        solid,
        draw=black!65,
        boxplot/every box/.style={
            fill=#7,
            draw=black!65,
            solid,
            line width=0.65pt
        },
        boxplot/every median/.style={
            draw=black,
            solid,
            line width=1pt
        },
        boxplot/every whisker/.style={
            draw=black!65,
            solid,
            line width=0.65pt
        }
    ] coordinates {};
}

\begin{tikzpicture}
\begin{axis}[
    width=\textwidth,
    height=4.3cm,
    xmin=0.55, xmax=14.75,
    ymin=0, ymax=1.0,
    ylabel={JSD},
    ylabel style={font=\scriptsize},
    ytick={0,0.2,0.4,0.6,0.8,1.0},
    yticklabel style={font=\scriptsize},
    xticklabel style={
        font=\scriptsize,
        rotate=35,
        anchor=east
    },
    tick label style={font=\scriptsize},
    ymajorgrids=true,
    grid style={dashed, gray!20},
    axis x line*=bottom,
    axis y line*=left,
    axis line style={black!65},
    tick style={black!65},
    boxplot/draw direction=y,
    boxplot/box extend=0.24,
    xtick={
        1.16,2.16,3.16,4.16,
        6.16,7.16,8.16,9.16,10.16,
        11.16,12.16,13.16,14.16
    },
    xticklabels={
        Airport,
        Cora,
        Disease,
        Citeseer,
        Cora,
        DBLP-A,
        Cora-CA,
        IMDB,
        Pubmed,
        DBLP-P,
        Aminer,
        MN-40,
        Pubmed
    },
    legend style={
        at={(0.5,1.15)},
        anchor=south,
        legend columns=4,
        draw=none,
        font=\scriptsize,
        /tikz/every even column/.append style={
            column sep=0.3cm
        }
    },
    legend cell align=left,
    clip=false
]

% ================= Legend =================

\addlegendimage{
    area legend,
    draw=black!65,
    fill=gNon
}
\addlegendentry{Graph non-bridge}

\addlegendimage{
    area legend,
    draw=black!65,
    fill=gBri
}
\addlegendentry{Graph bridge}

\addlegendimage{
    area legend,
    draw=black!65,
    fill=hHom
}
\addlegendentry{Hypergraph non-bridge}

\addlegendimage{
    area legend,
    draw=black!65,
    fill=hMix
}
\addlegendentry{Hypergraph bridge}

% ==================================================
%                    Graph datasets
% ==================================================

% ---------------- Airport ----------------
\mybox{1.00}
{0.000000}{0.009450}{0.034843}
{0.068188}{0.395816}{gNon}

\mybox{1.32}
{0.005050}{0.062305}{0.095579}
{0.137932}{0.276942}{gBri}

% ---------------- Cora ----------------
\mybox{2.00}
{0.000000}{0.000000}{0.014238}
{0.055973}{0.395816}{gNon}

\mybox{2.32}
{0.027890}{0.132563}{0.179276}
{0.230359}{0.419165}{gBri}

% ---------------- Disease ----------------
\mybox{3.00}
{0.000000}{0.000000}{0.000000}
{0.000000}{0.190875}{gNon}

\mybox{3.32}
{0.101046}{0.107218}{0.134157}
{0.147873}{0.179095}{gBri}

% ---------------- Pubmed ----------------
\mybox{4.00}
{0.000000}{0.000000}{0.009203}
{0.042974}{0.311278}{gNon}

\mybox{4.32}
{0.008005}{0.079779}{0.116030}
{0.163363}{0.387038}{gBri}

% ==================================================
%                 Hypergraph datasets
% ==================================================

% ---------------- Citeseer ----------------
\mybox{7.00}
{0.000000}{0.000000}{0.000000}
{0.023611}{0.425284}{hHom}

\mybox{7.32}
{0.000000}{0.003981}{0.036869}
{0.213493}{0.846092}{hMix}

% ---------------- Cora ----------------
\mybox{8.00}
{0.000000}{0.000000}{0.047016}
{0.123108}{0.626383}{hHom}

\mybox{8.32}
{0.000000}{0.073104}{0.189421}
{0.347231}{0.781893}{hMix}

% ---------------- DBLP Copub ----------------
\mybox{11.00}
{0.000000}{0.000000}{0.000000}
{0.024228}{0.429563}{hHom}

\mybox{11.32}
{0.000236}{0.057565}{0.218684}
{0.384847}{0.653561}{hMix}

% ---------------- Cora Coauth ----------------
\mybox{9.00}
{0.000000}{0.000000}{0.000000}
{0.007624}{0.642691}{hHom}

\mybox{9.32}
{0.000000}{0.000676}{0.022259}
{0.201157}{0.830249}{hMix}

% ---------------- IMDB ----------------
\mybox{12.00}
{0.000000}{0.020715}{0.055173}
{0.114281}{0.609987}{hHom}

\mybox{12.32}
{0.000000}{0.018956}{0.061463}
{0.153147}{0.777132}{hMix}

% ---------------- Pubmed ----------------
\mybox{14.00}
{0.000000}{0.001775}{0.007112}
{0.026145}{0.657617}{hHom}

\mybox{14.32}
{0.000000}{0.006808}{0.030691}
{0.158938}{0.919250}{hMix}

% ---------------- DBLP Coauth ----------------
\mybox{10.00}
{0.000000}{0.000000}{0.000000}
{0.015565}{0.543303}{hHom}

\mybox{10.32}
{0.000000}{0.002568}{0.022821}
{0.134954}{0.891160}{hMix}

% ---------------- Aminer ----------------
\mybox{6.00}
{0.000000}{0.000000}{0.000000}
{0.000000}{0.475448}{hHom}

\mybox{6.32}
{0.000000}{0.000082}{0.005449}
{0.056989}{0.830687}{hMix}

% ---------------- ModelNet40 ----------------
\mybox{13.00}
{0.000000}{0.000000}{0.011458}
{0.065508}{0.712642}{hHom}

\mybox{13.32}
{0.000000}{0.055694}{0.163448}
{0.317937}{0.807167}{hMix}

% ==================================================
%                    Decorations
% ==================================================

% Graph / Hypergraph separator
\draw[
    dashed,
    black!45,
    line width=0.8pt
]
(axis cs:5.15,0) --
(axis cs:5.15,1.0);

% Section headers
\node[
    font=\bfseries\scriptsize,
    anchor=south
]
at (axis cs:2.65,1.025)
{Graph datasets};

\node[
    font=\bfseries\scriptsize,
    anchor=south
]
at (axis cs:10.15,1.025)
{Hypergraph datasets};

\end{axis}
\end{tikzpicture}

\vspace{-2mm}
\caption{
Jensen-Shannon divergence (JSD) of bridge and non-bridge nodes across graph and hypergraph datasets.
}
\label{f4}
\vspace{-8mm}

\end{figure*}
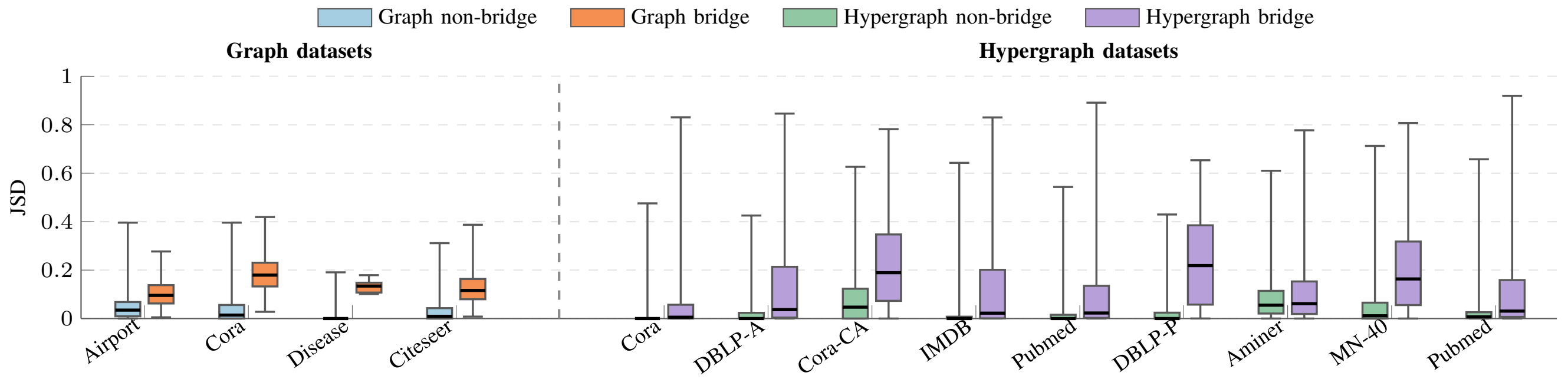

%% file: Figs/f5.tex
\begin{figure*}[t]
    \centering
    \includegraphics[
        width=\textwidth,
        keepaspectratio
    ]{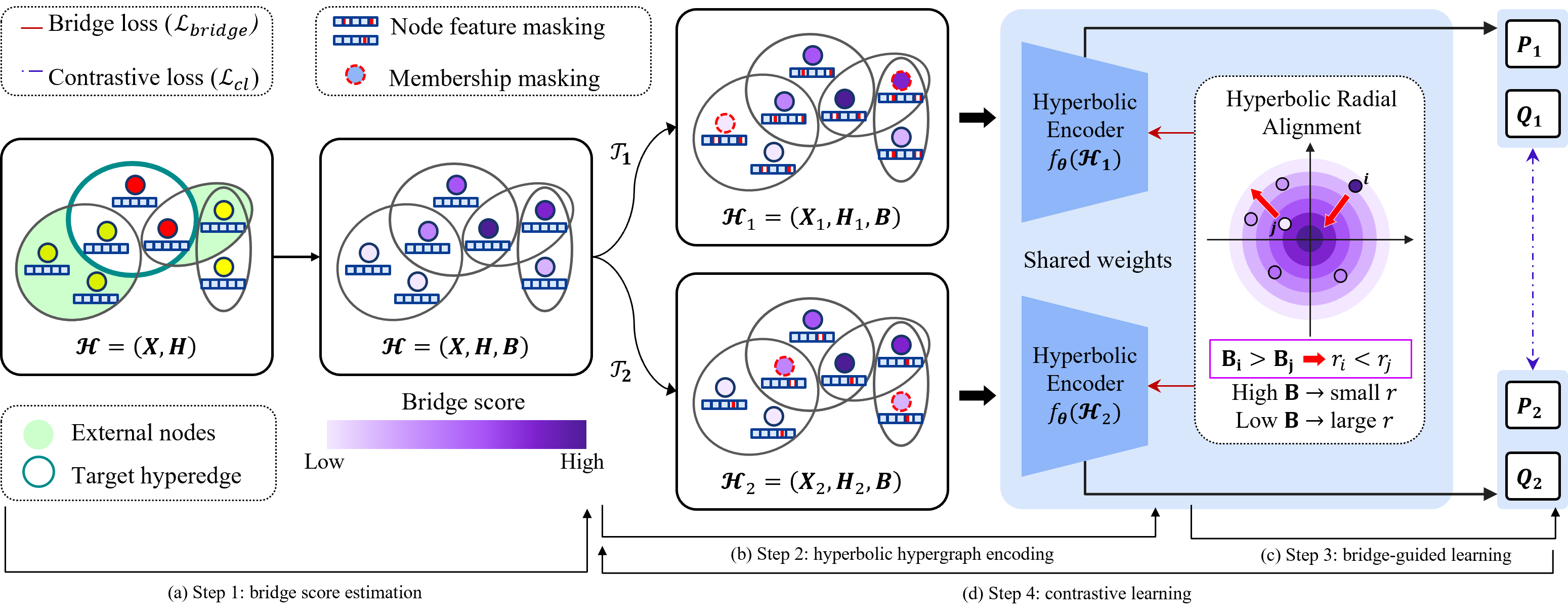}
    \vspace{-3mm}
    \caption{Overview of representation learning in \textsf{HyBridge}.}
    \label{f5}
    \vspace{-5mm}
\end{figure*}

%% file: s4.tex
\section{{\hbc}: The Proposed Framework}\label{s4}

\subsection{Notations and Overview}\label{4.1}
A hypergraph is defined as $\mathcal{H}=(V,E)$, where $V$ denotes the set of nodes and $E$ denotes the set of hyperedges.
Its structure is represented by an incidence matrix $\mathbf{H}\in\{0,1\}^{|V|\times|E|}$, where $\mathbf{H}(v,e)=1$ indicates that node $v$ is incident to hyperedge $e$.
Each hyperedge is associated with a positive weight, represented by a diagonal matrix $\mathbf{W}\in\mathbb{R}^{|E|\times|E|}$.
Node features are given by $\mathbf{X}\in\mathbb{R}^{|V|\times F}$, and we equivalently represent the hypergraph as $\mathcal{H}=(\mathbf{X},\mathbf{H})$.
The node and hyperedge degree matrices are denoted by $\mathbf{D}_V$ and $\mathbf{D}_E$, respectively.
In this paper, we address HRL under a \textit{fully unsupervised setting}, where the goal is to learn a hypergraph encoder $f_\theta$ that maps each node $v_i\in V$ to an $F'$-dimensional representation $\mathbf{p}_i\in\mathbb{R}^{F'}$.
The learned representations aim to preserve both the hypergraph structure and node attribute information in the latent space.

Fig.~\ref{f5} illustrates the overall HB-HRL process of {\hbc}, which consists of four main steps:
(a) \textit{bridge score estimation}, (b) \textit{hyperbolic hypergraph encoding}, (c) \textit{bridge-guided learning}, and (d) \textit{contrastive learning}.
Given a hypergraph $\mathcal{H}$ as input, \textit{bridge score estimation} first estimates the bridge score of each node (Fig.~\ref{f5}(a)).
Then, \textit{hyperbolic hypergraph encoding} maps nodes and hyperedges to representations in an $F'$-dimensional hyperbolic latent space (Fig.~\ref{f5}(b)).
Next, \textit{bridge-guided learning} aligns the relative ordering of node embedding radii with their bridge scores (Fig.~\ref{f5}(c)).
Finally, \textit{contrastive learning} preserves representation discriminability by treating the same node or hyperedge across two augmented views as a positive pair and different nodes or hyperedges as negative pairs (Fig.~\ref{f5}(d)).
\vspace{-2mm}
\subsection{Bridge Score Estimation}\label{4.2}
This step aims to quantify how strongly a node serves as a bridge across different communities in a hypergraph.
To this end, we estimate a bridge score for each hyperedge $e_j$ based on the label diversity of its external nodes $\mathcal{V}_\text{ext}(e_j)$, which belong to neighboring hyperedges but not to the target hyperedge $e_j$ itself.
We then obtain each node's bridge score by averaging the bridge scores of its incident hyperedges.

To estimate bridge scores without relying on ground-truth node labels, we first generate pseudo-cluster labels based solely on the hypergraph structure.
We apply METIS~\cite{karypis98:sisc}, which has been used in previous hypergraph studies~\cite{du22:bd}, to partition nodes into $K$ clusters based on structural connectivity.\footnote{We also evaluated representative community detection algorithms, including Louvain~\cite{blondel08:jstat} and Infomap~\cite{rosvall08:pnas}, and observed no substantial performance differences.}
Following~\cite{du22:bd}, we set $K$ to the number of ground-truth classes in each hypergraph, while using no ground-truth label assignments.
Using the resulting pseudo-cluster labels, we then characterize the external nodes $\mathcal{V}_\text{ext}(e_j)$ for each hyperedge $e_j$ using two hyperedge-level metrics adopted in previous hypergraph studies: label entropy~\cite{chun25:icdm} and label homophily~\cite{li25:aaai}.

Unlike these studies, which compute these metrics using the labels of nodes within the target hyperedge, we compute both metrics using the pseudo-cluster labels of external nodes.\footnote{A performance comparison with the approach using nodes within the target hyperedge is presented in Section~\ref{5.2}-EQ3.}
We then combine the \textit{normalized label entropy} ($\mathrm{Ent}_\text{ext}(e_j)$) and the \textit{complement of label homophily} ($1-\mathrm{Hom}_\text{ext}(e_j)$) with equal weights to obtain the bridge score of each hyperedge.
Finally, we compute each node's bridge score by averaging the bridge scores of its incident hyperedges.

\subsection{Hyperbolic Hypergraph Encoding}\label{4.3}
This step aims to learn node representations that capture groupwise relationships in hyperbolic space.
To this end, {\hbc} employs HB-HNHN, a hyperbolic encoder that extends the node-to-hyperedge and hyperedge-to-node aggregation operations of the Euclidean HNHN~\cite{don20:icmlw} to hyperbolic space.
We choose HNHN as the base encoder because its two-stage message passing directly operates on hyperedges, thereby preserving groupwise structural information that may be lost when hyperedges are reduced to pairwise connections.

Existing HB-GRL methods typically perform linear transformations and aggregation operations in tangent spaces.
Specifically, hyperbolic representations are projected onto tangent spaces via logarithmic maps $\text{log}_o(\cdot)$, where standard Euclidean operations can be performed, and then mapped back to hyperbolic space via exponential maps $\text{exp}_o(\cdot)$.
Following this approach, HB-HNHN first maps the initial Euclidean node features to hyperbolic space via an exponential map.
It then performs node-to-hyperedge and hyperedge-to-node aggregation in the tangent space and maps the resulting representations back to hyperbolic space, as formulated in Eq.~\ref{eq:hgconv}.

\small
\begin{equation}
\begin{aligned}
\mathbf{Q}^k &=
\exp_o^{c^k}
\left(
\sigma\left(
\mathbf{D}_E^{-1}\mathbf{H}^\top
\log_o^{c^k}(\mathbf{P}^k)
\mathbf{\Theta}_N^k
+
b_N^k
\right)
\right), \\
\mathbf{P}^{k+1} &=
\exp_o^{c^{k+1}}
\left(
\sigma\left(
\mathbf{D}_V^{-1}\mathbf{H}
\log_o^{c^k}(\mathbf{Q}^k)
\mathbf{\Theta}_E^k
+
b_E^k
\right)
\right),
\end{aligned}
\label{eq:hgconv}
\end{equation}
\normalsize
where $\mathbf{P}^k$ and $\mathbf{Q}^k$ denote the node and hyperedge embeddings at the $k$-th layer, respectively; $\mathbf{H}$ is the incidence matrix; $\mathbf{D}_V$ and $\mathbf{D}_E$ are the node and hyperedge degree matrices, respectively; $\mathbf{\Theta}_N^k$ and $\mathbf{\Theta}_E^k$ denote learnable transformation matrices; $b_N^k$ and $b_E^k$ denote learnable bias terms; and $\sigma(\cdot)$ denotes a nonlinear activation function.

The node embeddings $\mathbf{P}^k$ are first mapped to the tangent space, where the representations of incident nodes are aggregated for each hyperedge via $\mathbf{H}^{\top}$.
The aggregated representations are then transformed using the learnable parameters $\mathbf{\Theta}_N^k$ and $b_N^k$ and mapped back to hyperbolic space, yielding the hyperedge embeddings $\mathbf{Q}^k$.
Similarly, the hyperedge embeddings $\mathbf{Q}^k$ are mapped to the tangent space and aggregated for each incident node via $\mathbf{H}$.
The aggregated representations are then transformed using $\mathbf{\Theta}_E^k$ and $b_E^k$ and mapped back to hyperbolic space, yielding the next-layer node embeddings $\mathbf{P}^{k+1}$.
% \vspace{-3mm}

\subsection{Bridge-guided Learning}\label{4.4}
This step aims to determine the appropriate embedding radii of nodes within the hierarchical center--periphery structure using their bridge scores as structural guidance.
Accordingly, we encourage nodes with higher bridge scores to be positioned closer to the hyperbolic center.

We first obtain the node embeddings through the hyperbolic hypergraph encoder $f_\theta$.
Given the embedding $\mathbf{p}_i=f_\theta(v_i)$ of node $v_i$, its hyperbolic radius is defined as $r_i=d_{\mathbb{H}}(\mathbf{p}_i,\mathbf{o})$, where $d_{\mathbb{H}}(\cdot,\cdot)$ denotes the hyperbolic distance function and $\mathbf{o}$ denotes the origin of hyperbolic space.
We then randomly sample a set of node pairs $\mathcal{P}$ and construct $\mathcal{P}^{+}$ based on their bridge-score ordering.
Specifically, if node $v_i$ has a higher bridge score than node $v_j$ ($B_i>B_j$) but is positioned farther from the hyperbolic center ($r_i>r_j$), a penalty is imposed to encourage $r_i<r_j$.
Based on this ranking principle, we define the bridge-guided loss as follows:

\small
\begin{equation}
\mathcal{L}_{\mathrm{bridge}}
=
\frac{1}{|\mathcal{P}^{+}|}
\sum_{(i,j)\in\mathcal{P}^{+}}
\max(0, r_i-r_j),
\label{eq:bridge_loss}
\end{equation}
\normalsize
where $\mathcal{P}$ denotes the set of randomly sampled node pairs, and $\mathcal{P}^{+}=\{(i,j)\in\mathcal{P}\mid B_i>B_j\}$ denotes the subset of pairs in which node $v_i$ has a higher bridge score than node $v_j$.

% \vspace{-2mm}
\subsection{Contrastive Learning}\label{4.5}
This step aims to preserve representation discriminability among nodes while maintaining the radial structure guided by their bridge scores.
Since nodes with similar bridge scores may be positioned at similar radii, we employ contrastive learning to distinguish their representations.
Specifically, contrastive learning encourages representations of the same node across two augmented views to be closer while pushing representations of different nodes farther apart.

We first generate two views of the original hypergraph by randomly masking node features and node--hyperedge memberships according to their respective augmentation ratios, where higher ratios indicate a larger proportion of masked elements.
We then obtain node embeddings $\mathbf{P}$ and hyperedge embeddings $\mathbf{Q}$ for each view using the same hyperbolic hypergraph encoder $f_{\theta}$.
Next, we perform contrastive learning at both the \textit{node level}, which aligns representations of the same node across the two views, and the \textit{hyperedge level}, which aligns representations of the same hyperedge across the two views.

More specifically, for node-level contrastive learning, {\hbc} uses the representation of node $v_i$ from the first view, $\mathbf{z}_{1,i}$, as the anchor, its counterpart from the second view, $\mathbf{z}_{2,i}$, as the positive sample, and all other node representations $\mathbf{z}_{2,k}$, where $k \neq i$, as negative samples.
The contrastive loss for each node $v_i$ is defined as follows:

\small
\begin{equation}
\ell_n(\mathbf{z}_{1,i}, \mathbf{z}_{2,i}) = - \log \frac{\exp\left(s(\mathbf{z}_{1,i}, \mathbf{z}_{2,i}) / \tau_n\right)}{\sum_{k=1}^{|V|} \exp\left(s(\mathbf{z}_{1,i}, \mathbf{z}_{2,k}) / \tau_n\right)},
\end{equation}
\normalsize
where $\tau_n$ is a temperature parameter~\cite{lee23:aaai}, and $s(\cdot, \cdot)$ is a scoring function based on hyperbolic distance that assigns scores between 0 and 1, with higher scores indicating closer representations in hyperbolic space~\cite{zhang25:kdd}.
The overall node-level contrastive loss is then computed by averaging the losses over all nodes in both directions between the two views:

\vspace{-2mm}
\small
\begin{equation}\label{ln}
\mathcal{L}_{n} = \frac{1}{2|V|} \sum_{i=1}^{|V|} \left\{ \ell_n(\mathbf{z}_{1,i}, \mathbf{z}_{2,i}) + \ell_n(\mathbf{z}_{2,i}, \mathbf{z}_{1,i}) \right\}.
\end{equation}
\normalsize

For hyperedge-level contrastive learning, {\hbc} uses the representation of hyperedge $e_j$ from the first view, $\mathbf{y}_{1,j}$, as the anchor, its counterpart from the second view, $\mathbf{y}_{2,j}$, as the positive sample, and all other hyperedge representations $\mathbf{y}_{2,k}$, where $k \neq j$, as negative samples.
The contrastive loss for each hyperedge $e_j$ is defined as follows:

\vspace{-2mm}
\small
\begin{equation}
\ell_h(\mathbf{y}_{1,j}, \mathbf{y}_{2,j}) = - \log \frac{\exp\left(s(\mathbf{y}_{1,j}, \mathbf{y}_{2,j}) / \tau_h\right)}{\sum_{k=1}^{|E|} \exp\left(s(\mathbf{y}_{1,j}, \mathbf{y}_{2,k}) / \tau_h\right)},
\end{equation}
\normalsize
where $\tau_h$ is a temperature parameter~\cite{lee23:aaai}.
The overall hyperedge-level contrastive loss is then computed by averaging the losses over all hyperedges in both directions between the two views:

\vspace{-2mm}
\small
\begin{equation}\label{lh}
\mathcal{L}_{h} = \frac{1}{2|E|} \sum_{j=1}^{|E|} \left\{ \ell_h(\mathbf{y}_{1,j}, \mathbf{y}_{2,j}) + \ell_h(\mathbf{y}_{2,j}, \mathbf{y}_{1,j}) \right\}.
\end{equation}
\normalsize

By combining the node-level and hyperedge-level contrastive losses in Eqs.~\ref{ln} and~\ref{lh}, respectively, the overall contrastive loss is defined as follows:

\vspace{-2mm}
\small
\begin{equation}\label{final_lh}
\mathcal{L}_{cl}=\omega_{n}\mathcal{L}_{\textit{n}} + \omega_{h}\mathcal{L}_{\textit{h}}.
\end{equation}
\normalsize

Finally, the overall objective of {\hbc} jointly optimizes the bridge-guided radial organization and representation discriminability:

\vspace{-2mm}
\small
\begin{equation}\label{final_loss}
\mathcal{L}=\omega_{bridge}\mathcal{L}_{\textit{bridge}} + \mathcal{L}_{\textit{cl}}.
\end{equation}
\normalsize

Through this joint optimization, {\hbc} learns discriminative node representations while preserving the bridge-guided hierarchical center--periphery structure in hyperbolic space.

\subsection{Time Complexity Analysis}\label{4.6}
We summarize the overall learning process of {\hbc} in Algorithm~\ref{a1}.
{\hbc} first initializes the model parameters $\theta$ (line 1) and estimates node-level bridge scores by computing the bridge score of each hyperedge and averaging the scores of its incident hyperedges for each node (lines 2--6).
During training, {\hbc} generates two augmented hypergraph views and obtains their node and hyperedge embeddings using the hyperbolic hypergraph encoder (lines 7--11).
It then randomly samples node pairs, constructs $\mathcal{P}^{+}$ according to their bridge-score ordering, and computes the bridge-guided loss (lines 12--14), followed by the contrastive loss (line 15).
Finally, {\hbc} jointly optimizes the two losses to update the model parameters (lines 16--17) and generates the final node representations (lines 19--20).

We further analyze the time complexity of {\hbc}.
Step 1 is a one-time preprocessing step that generates pseudo-cluster labels, identifies external node sets, and computes node bridge scores.
Step 2 requires $O(L(|V|+|E|+S))$ per epoch, where $L$ is the number of encoder layers and $S$ is the number of node–hyperedge incidences.
Step 3 requires $O(|\mathcal{P}|)$ for sampled node pairs $\mathcal{P}$.
Step 4 requires $O(|V|^2+|E|^2)$ for node- and hyperedge-level contrastive objectives.
Therefore, the overall per-epoch time complexity of {\hbc} is $O(L(|V|+|E|+S)+|\mathcal{P}|+|V|^2+|E|^2)$, excluding the one-time preprocessing cost. 
We further evaluate its practical training efficiency and preprocessing overhead in Section~\ref{5.2}-EQ6.

\input{algorithm/a1}

%% file: algorithm/a1.tex
\begin{algorithm}[t]
\scriptsize
\def\arraystretch{1.25}
\caption{\textsf{HyBridge} framework}
\label{a1}

  \begin{flushleft}
      \textbf{Input:} Hypergraph H = (X, H) (node vector matrix X, incidence matrix H)\\
      \textbf{Output:} Learned node representations P
\end{flushleft}

\begin{algorithmic}[1]

    \State Initialize model parameters $\theta$
    
    % ===== Step 1: Bridge Score Estimation =====
    \State Assign pseudo cluster label using METIS
    \Comment{\textcolor{blue}{Step 1: Bridge score estimation}}
    
    \State Identify external node sets
    $\{\mathcal{V}_{\mathrm{ext}}(e_j)\}_{e_j \in E}$

    \State Compute $\mathbf{Ent}_{\mathrm{ext}}(e_j)$ and
    $\mathbf{Hom}_{\mathrm{ext}}(e_j)$ using pseudo cluster label

    \State Hyperedge bridge score 
    $
    \leftarrow
    \frac{1}{2}\left(
    \mathbf{Ent}_{\mathrm{ext}}
    +\mathbf{1}
    -\mathbf{Hom}_{\mathrm{ext}}
    \right)$

    \State Compute node-level bridge scores
    $\mathbf{B}$ by averaging incident
    hyperedge bridge scores

    % ===== Representation Learning =====
    \For{epoch $\leftarrow$ 1 \text{to epochs}}
    \Comment{\textcolor{blue}{Step 2: Hyperbolic hypergraph encoding}}
        \State Generate two augmented views
        $\mathcal{H}^{(1)},\mathcal{H}^{(2)}$
        
        \For{$a \in \{1,2\}$}

            \State $(\mathbf{P}^{(a)},\mathbf{Q}^{(a)})
            \leftarrow f_{\theta}(\mathcal{H}^{(a)})$
            via Eq.~\ref{eq:hgconv}

        \EndFor

        % ===== Step 2: Bridge-guided Learning =====
        \State Sample ordered node pairs $\mathcal{P}$
        \Comment{\textcolor{blue}{Step 3: Bridge-guided learning}}
        \State Construct $\mathcal{P}^{+}$ based on $\mathbf{B}$

        \State Compute bridge-guided loss
        $\mathcal{L}_{\mathrm{bridge}}$
        via Eq.~\ref{eq:bridge_loss}

        % ===== Step 3: Contrastive Learning =====
        \State Compute contrastive loss $\mathcal{L}_{cl}$ via Eq.~\ref{final_lh}
        \Comment{\textcolor{blue}{Step 4: Contrastive learning}}

        % ===== Joint Optimization =====
        \State $\mathcal{L}
        \leftarrow
        \omega_{bridge}\mathcal{L}_{\mathrm{bridge}}
        +\mathcal{L}_{cl}$

        \State Update $\theta$ by minimizing $\mathcal{L}$

    \EndFor

    \State $(\mathbf{P},\mathbf{Q})
    \leftarrow f_{\theta}(\mathcal{H})$

    \State \Return $\mathbf{P}$

\end{algorithmic}
\end{algorithm}

\setlength{\textfloatsep}{0.1pt}

%% file: s5.tex
\section{Evaluation}\label{s5}
In this section, we validate the effectiveness of {\hbc} through extensive experiments designed to answer the following evaluation questions (EQs):
\begin{itemize}[leftmargin=*]
    \item \textbf{EQ1}: What are the contributions of hyperbolic representation learning and bridge-guided learning?
    \item \textbf{EQ2}: Does {\hbc} outperform existing state-of-the-art methods across different downstream tasks?
    \item \textbf{EQ3}: How does using within-hyperedge versus external-node label distributions for bridge score estimation affect the performance of {\hbc}?
    \item \textbf{EQ4}: Does bridge-guided learning induce consistent structural patterns across hyperbolic radii?
    \item \textbf{EQ5}: How sensitive is {\hbc} to its key hyperparameters?
    \item \textbf{EQ6}: How computationally efficient and scalable is {\hbc}?
\end{itemize}

\subsection{Experimental Settings}\label{5.1}

\noindent \textbf{Datasets.}
We use 10 real-world hypergraph datasets from diverse domains, including co-citation (\ie, Cora~\cite{yad19:neurips}, Citeseer~\cite{yad19:neurips}, and PubMed~\cite{yad19:neurips}), co-authorship (\ie, Cora-CA~\cite{yad19:neurips}, DBLP-P~\cite{yad19:neurips}, DBLP-A~\cite{wan19:www}, and AMiner~\cite{zha19:kdd}), computer graphics (\ie, MN-40~\cite{wu15:cvpr}), movie--actor relationships (\ie, IMDB~\cite{wan19:www}), and news content (\ie, 20News~\cite{kel17}).
These datasets constitute the union of those adopted in prior work~\cite{lee23:aaai,kim24:iclr}.
Table~\ref{tab_4} summarizes their basic statistics.

\input{Tabs/t4}

\vspace{1mm}
\noindent \textbf{Competing Methods.}
We compare {\hbc} with 22 competing methods.
The Euclidean methods include (1) 15 HRL methods: HGNN~\cite{fen19:aaai}, HyperGCN~\cite{yad19:neurips}, HNHN~\cite{don20:icmlw}, UniGNN variants (\ie, UniGCN, UniGIN, and UniGCNII)~\cite{hua21:ijcai}, AllSet~\cite{chi21:iclr}, ED-HNN~\cite{wan23:iclr}, PhenomNN~\cite{wan23:icml}, TriCL~\cite{lee23:aaai}, HyperGCL~\cite{wei22:nips}, HyperGRL~\cite{du22:bd}, HypeBoy~\cite{kim24:iclr}, SE-HSSL~\cite{li24:ijcai}, and VilLain~\cite{lee24:web};
(2) three GRL methods: GGD~\cite{zhe22:nips}, GraphMAE2~\cite{hou23:www}, and MaskGAE~\cite{li23:kdd};
and (3) one baseline method: MLP~\cite{ga98:ae}.
The hyperbolic methods include (1) two HB-GRL methods: HB-GCN~\cite{chami19:nips} and HB-GCL~\cite{zhang25:kdd}; and (2) one HB-HRL method: THINK~\cite{agarwal22:icdm}.

Since THINK is designed for temporal hypergraphs, we use its static variant without temporal information.
For methods originally designed for graphs, we follow the experimental settings of prior work~\cite{kim24:iclr}.
Specifically, for GGD, we replace its original feature and topology augmentations with the augmentation strategy used in HypeBoy~\cite{kim24:iclr}.
For GraphMAE2, MaskGAE, HB-GCN, and HB-GCL, we first convert each hypergraph into a graph via clique expansion and then apply the respective methods.

\vspace{1mm}
\noindent \textbf{Target downstream tasks and evaluation protocols.}
To evaluate {\hbc} across different downstream tasks, we consider node classification~\cite{wei22:nips, kim24:iclr} and hyperedge prediction~\cite{lee23:aaai, kim24:iclr}.
For node classification, following~\cite{wei22:nips, kim24:iclr}, we randomly split nodes into training, validation, and test sets with ratios of 1\%, 1\%, and 98\%, respectively, and report the mean and standard deviation of accuracy over 20 random splits.
Each self-supervised method first learns node representations, after which an MLP classifier is trained on the learned representations~\cite{kim24:iclr}.

For hyperedge prediction, following~\cite{wei22:nips, kim24:iclr}, we randomly split hyperedges into training, validation, and test sets with ratios of 60\%, 20\%, and 20\%, respectively, and report the average AUROC over 20 random splits.
Negative hyperedges are generated using Size-Negative-Sampling~\cite{pat20:pakdd}.
For Euclidean-based methods, hyperedge representations are obtained by applying max-min pooling to the embeddings of nodes contained in each hyperedge~\cite{yad19:neurips, kim24:iclr}.
For hyperbolic-based methods, standard Euclidean pooling may not adequately characterize the spatial distribution of node embeddings within each hyperedge.
Therefore, we incorporate the \textit{mean}, \textit{maximum}, and \textit{standard deviation} of pairwise hyperbolic distances among nodes within each hyperedge, along with the logarithm of the hyperedge size, as additional hyperedge-level features.\footnote{We also apply the same statistical features to Euclidean-based methods; however, we observe no significant performance difference compared with their original counterparts, suggesting that max-min pooling already captures sufficient information for hyperedge representation in Euclidean space.}

\vspace{1mm}
\noindent \textbf{Reproducibility.}
We conduct all experiments on a Linux server equipped with an AMD 7960X processor, a 2TB NVMe SSD, 128GB of DDR5 memory, and an NVIDIA RTX A6000 GPU with 48GB of memory.
For the competing methods, we use the source code provided by the authors of the original papers~\cite{fen19:aaai, yad19:neurips, don20:icmlw, hua21:ijcai, chi21:iclr, wan23:iclr, wan23:icml, lee23:aaai, wei22:nips, du22:bd, kim24:iclr, li24:ijcai, lee24:web, zhe22:nips, hou23:www, li23:kdd, ga98:ae, liu19:nips, zhang25:kdd, agarwal22:icdm}.
For each method, we use the best hyperparameter setting identified through grid search over the ranges suggested in the corresponding paper.

We tune the hyperparameters of {\hbc} via grid search over the following ranges.
The loss weights $\omega_n$ and $\omega_h$ are selected from $\{2^{-4}, 2^{-3}, \dots, 2^{4}\}$, while $\omega_{\mathrm{bridge}}$ is selected from $\{10^{-4}, 5 \times 10^{-4}, 10^{-3}, 5 \times 10^{-3}, \dots, 1\}$.
The temperature parameters for the contrastive losses, $\tau_n$ and $\tau_h$, are selected from $\{0.1, 0.2, \dots, 1.0\}$.
For hypergraph augmentation, the feature, incidence, and hyperedge masking rates are varied within $\{0.1, 0.2, \dots, 0.9\}$.
The hidden dimension is selected from $\{2^6, 2^7, \dots, 2^{10}\}$, and the learning rate is selected from $\{10^{-5}, 5 \times 10^{-5}, 10^{-4}, 5 \times 10^{-4}, 10^{-3}\}$.
The weight decay is tuned over $\{10^{-5}, 5 \times 10^{-5}, 10^{-4}, 5 \times 10^{-4}\}$.
The number of sampled node pairs $|\mathcal{P}|$ is selected from $\{2^{12}, 2^{13}, \dots, 2^{17}\}$.
For reproducibility, we provide the code at https://anonymous.4open.science/r/HyBridge/.

\subsection{Experimental Results}\label{5.2}

\noindent \textbf{EQ1: Ablation study.}
We conduct an ablation study with three variants: \textbf{V1} performs contrastive learning in Euclidean space; \textbf{V2} performs contrastive learning in hyperbolic space; and \textbf{V3} additionally incorporates bridge-guided learning in hyperbolic space.
Thus, comparing \textbf{V1} and \textbf{V2} evaluates the effect of hyperbolic geometry, while comparing \textbf{V2} and \textbf{V3} assesses the additional contribution of bridge-guided learning.

\input{Tabs/t5}

Table~\ref{tab_5} presents the results.
For each dataset, the best and second-best performances are marked in \hly{\textbf{bold}} and \hlb{\mbox{\underline{underline}}}, respectively.
\textit{AR} denotes the average rank.
We make two key observations.
First, \textbf{V2} outperforms \textbf{V1} on some datasets but shows comparable or lower performance on others, indicating that hyperbolic geometry alone does not consistently improve hypergraph representation learning.
Second, \textbf{V3} achieves the best average rank, demonstrating that incorporating bridge-guided learning consistently improves the effectiveness of hyperbolic representations.
These results show that hyperbolic geometry alone provides mixed benefits, whereas bridge guidance provides effective structural guidance for learning hyperbolic hypergraph representations.

\vspace{1mm}
\noindent \textbf{EQ2: Performance comparison.}
We evaluate the performance of {\hbc} against the 22 competing methods introduced in Section~\ref{5.1}.
Table~\ref{tab_6} presents the node classification results.
OOT (out of time) indicates that a method did not complete within 24 hours, while OOM (out of memory) indicates that a method exceeded the memory capacity of the 48GB GPU.
We make three key observations.
First, {\hbc} consistently outperforms all competing methods, achieving the highest accuracy across all datasets and the lowest average rank (AR) of 1.0.
Second, existing hyperbolic methods, including HB-GCN (AR: 13.2) and HB-GCL (AR: 13.6), generally perform worse than strong Euclidean HRL methods, such as HypeBoy (AR: 4.2) and TriCL (AR: 5.2).
Third, THINK, which is originally designed for temporal hypergraphs, achieves an AR of 12.2 when evaluated using its static variant.
Similar trends are observed for hyperedge prediction, as reported in Table~\ref{tab_7}.

\input{Tabs/t6}
\input{Tabs/t7}

\vspace{1mm}
\noindent \textbf{EQ3: Bridge score design choice validation.}
We investigate the impact of using external-node label distributions, rather than within-hyperedge label distributions, for bridge score estimation in {\hbc}.
Specifically, we compare bridge scores estimated using the label distribution of nodes within the target hyperedge with those estimated using the label distribution of external nodes, as proposed in {\hbc}.
We vary only the node set used for bridge score estimation while keeping all other components and training settings unchanged.
We evaluate the learned node representations on node classification.
Table~\ref{tab_8} presents the node classification results.
The proposed approach using external-node label distributions achieves higher node classification accuracy across all datasets.
These results suggest that using external-node label distributions enables {\hbc} to better capture the bridging roles of individual hyperedges, thereby providing more effective structural guidance for hyperbolic representation learning.

\input{Tabs/t8}

\vspace{1mm}
\noindent \textbf{EQ4: Bridge-guided embedding radius analysis.}
As discussed in Section~\ref{s3.4}, HB-HRL does not exhibit consistent relationships between node radial positions and neighborhood label diversity.
To investigate whether bridge-guided learning addresses this limitation, we examine the radial patterns of the node representations learned by {\hbc} in terms of label entropy and homophily.
Specifically, following the same procedure as in Section~\ref{s3.4}, we divide nodes into ten groups according to their radial positions and compute the average neighborhood label entropy and homophily within each group.

Fig.~\ref{f6} presents the radial patterns of label entropy and homophily, where the \textit{x}-axis represents the ten node groups ordered from the center to the periphery and the \textit{y}-axis represents the average label entropy or homophily within each group.
Overall, label entropy tends to decrease while homophily tends to increase from the center toward the periphery.
Despite some fluctuations across datasets, these trends indicate that bridge-guided learning encourages a more consistent center--periphery organization, positioning nodes with more diverse neighborhood label distributions closer to the center.

\input{Figs/f6}
\input{Figs/f8}

\vspace{1mm}
\noindent \textbf{EQ5: Parameter sensitivity.}
To investigate the \textit{learning stability} of {\hbc}, we examine its sensitivity to key hyperparameters associated with (1) bridge-guided learning, including the bridge loss weight $\omega_{\mathrm{bridge}}$ and the number of sampled node pairs $|\mathcal{P}|$, and (2) contrastive learning, including the augmentation ratios for feature masking and membership masking.
For bridge-guided learning, we select $\omega_{\mathrm{bridge}}$ and $|\mathcal{P}|$ from the sets $\{10^{-4}, 5 \times 10^{-4}, 10^{-3}, 5 \times 10^{-3}, \dots, 1\}$ and $\{2^{12}, 2^{13}, \dots, 2^{17}\}$, respectively.

Fig.~\ref{f8} presents the node classification accuracy under different bridge loss weights on (a) Citeseer and (b) Cora, where the \textit{x}-axis represents $\omega_{\mathrm{bridge}}$ and the \textit{y}-axis represents node classification accuracy.
As $\omega_{\mathrm{bridge}}$ increases, performance initially improves but begins to decline beyond a certain point.
This indicates that an appropriate bridge loss weight is important for effective representation learning, while excessive emphasis on bridge-guided learning can adversely affect performance.

Fig.~\ref{f10} presents the node classification accuracy under different numbers of sampled node pairs on (a) Citeseer and (b) Cora, where the \textit{x}-axis represents $|\mathcal{P}|$ and the \textit{y}-axis represents node classification accuracy.
Performance remains relatively stable once a sufficient number of node pairs is sampled.
This suggests that {\hbc} is relatively insensitive to $|\mathcal{P}|$ beyond this point.

\input{Figs/f10}

For contrastive learning, following~\cite{lee23:aaai}, we select the feature masking and membership masking ratios from the set $\{0.1, 0.2, \ldots, 0.9\}$.
Fig.~\ref{f9} presents the differences in node classification accuracy between HyBridge and the second-best competing model Hypeboy under different combinations of augmentation ratios on (a) Citeseer and (b) Cora, where the \textit{x}-axis and \textit{y}-axis represent the feature masking and membership masking ratios, respectively.
We make two key observations.
First, excessively high feature masking ratios (0.8--0.9) lead to substantial performance degradation on both datasets, particularly on Cora, indicating that excessive feature removal can negatively affect representation learning.
Second, the favorable feature masking ratios differ across datasets: Citeseer achieves better performance with ratios of 0.5--0.7, whereas Cora performs better with ratios of 0.2--0.3.
These results highlight the importance of selecting appropriate augmentation ratios according to dataset characteristics.

\input{Figs/f9}
\input{Figs/f12}
\input{Figs/f13}
\input{Figs/f11}

\vspace{1mm}
\noindent \textbf{EQ6: Computational efficiency and scalability.}
To assess the computational efficiency of {\hbc}, we evaluate both the preprocessing overhead of Step 1 (bridge score estimation) and the training efficiency of the remaining steps.
First, we measure the total preprocessing time, including pseudo-cluster label generation using METIS and node-level bridge score estimation.
We evaluate the preprocessing time on Citeseer and Cora, as well as the two largest datasets, DBLP-P and AMiner.
Fig.~\ref{f12} presents the total preprocessing time, where the \textit{x}-axis represents the datasets and the \textit{y}-axis represents the preprocessing time.
The total preprocessing time is less than one second on three datasets and reaches only about two seconds on DBLP-P, demonstrating the low computational overhead of the preprocessing steps.

Second, we compare the training efficiency of {\hbc} with that of competing methods by measuring the average training time per epoch.
Fig.~\ref{f13} presents the relationship between training time and node classification accuracy on (a) Citeseer and (b) Cora, where the \textit{x}-axis represents the average training time per epoch and the \textit{y}-axis represents node classification accuracy.
Although {\hbc} is not the fastest method, it achieves a favorable trade-off between training time and classification accuracy.
In particular, {\hbc} requires less training time than existing contrastive HRL methods, such as TriCL and HyperGCL, while achieving higher accuracy on both datasets.

Finally, we empirically evaluate the scalability of {\hbc} as the size of the input hypergraph increases.
To this end, we train {\hbc} for 20 epochs on each dataset while varying the proportion of hyperedges used for training from 10\% to 100\% in increments of 10\%, and measure the average training time per epoch.
Fig.~\ref{f11} presents the results on the two largest datasets, (a) DBLP-P and (b) AMiner, where the \textit{x}-axis represents the proportion of hyperedges used for training and the \textit{y}-axis represents the average training time per epoch.
The training time increases approximately linearly with the proportion of hyperedges on both datasets, indicating approximately linear empirical scaling over the evaluated range.

%% file: Tabs/t4.tex
\begin{table*}[t]
\centering
\caption{Statistics of ten real-world hypergraphs used in our experiments}
\label{tab_4}
\scriptsize
\vspace{-2mm}
\def\arraystretch{1.25}

\resizebox{0.9\textwidth}{!}{%
\begin{tabular}{c||cccccccccc}
\toprule
  & \textbf{Citeseer}
  & \textbf{Cora}
  & \textbf{DBLP-A}
  & \textbf{Cora-CA}
  & \textbf{IMDB}
  & \textbf{Pubmed}
  & \textbf{DBLP-P}
  & \textbf{Aminer}
  & \textbf{MN-40}
  & \textbf{20News} \\
\midrule
\midrule

\# of Nodes
& 1,458 & 1,434 & 2,591 & 2,388
& 3,939 & 3,840 & 41,302 & 20,201
& 12,311 & 16,242 \\

\# of Hyperedges
& 1,079 & 1,579 & 2,690 & 1,072
& 2,015 & 7,963 & 22,263 & 8,052
& 12,311 & 100 \\

\# of Classes
& 6 & 7 & 4 & 7 & 3
& 3 & 6 & 12 & 40 & 4 \\

\# of Features
& 3,703 & 1,433 & 334 & 1,433
& 3,066 & 500 & 1,425 & 500
& 100 & 100 \\

\bottomrule
\end{tabular}%
}
\label{t4}
\vspace{-5mm}
\end{table*}

%% file: Tabs/t5.tex
\begin{table*}[t]
\centering
\caption{Ablation study: Accuracy comparison of 3 variants of {\hbc} with node classification on 10 real-world hypergraphs}
\label{tab_5}
\normalsize
% \vspace{-2mm}
\def\arraystretch{1.3}
\resizebox{\textwidth}{!}{
\begin{tabular}{c|cc||cccccccccc|c}
\toprule
  & \textbf{Hyperbolic} & \textbf{Bridge}
  & \textbf{Citeseer} & \textbf{Cora} & \textbf{DBLP-A}& \textbf{Cora-CA}
  & \textbf{IMDB} & \textbf{Pubmed} & \textbf{DBLP-P}& \textbf{AMiner}
  & \textbf{MN-40} & \textbf{20News} & \textbf{A.R.}  \\
\midrule
\midrule

\textbf{V1}
& \textcolor{red}{\xmark}
& \textcolor{red}{\xmark}
& $49.2 \pm 8.4$
& \hlb{\mbox{\underline{$65.8 \pm 6.2$}}}
& \hlb{\mbox{\underline{$79.4 \pm 2.7$}}}
& $66.7 \pm 5.5$
& $51.5 \pm 3.2$
& $68.3 \pm 5.2$
& $87.2 \pm 0.6$
& $34.8 \pm 2.5$
& \hlb{\mbox{\underline{$94.8 \pm 1.3$}}}
& $77.3 \pm 1.0$
& $2.7$ \\

\textbf{V2}
& \textcolor{teal}{\checkmark}
& \textcolor{red}{\xmark}
& \hlb{\mbox{\underline{$59.5 \pm 8.5$}}}
& $64.6 \pm 9.1$
& $78.8 \pm 4.4$
& \hlb{\mbox{\underline{$70.4 \pm 5.4$}}}
& \hlb{\mbox{\underline{$52.5 \pm 3.2$}}}
& \hlb{\mbox{\underline{$74.0 \pm 5.1$}}}
& \hlb{\mbox{\underline{$89.0 \pm 0.6$}}}
& \hlb{\mbox{\underline{$35.3 \pm 5.2$}}}
& $94.5 \pm 1.3$
& \hlb{\mbox{\underline{$78.9 \pm 0.5$}}}
& $2.3$ \\

\textbf{V3}
& \textcolor{teal}{\checkmark}
& \textcolor{teal}{\checkmark}

& \hly{\textbf{60.7 $\pm$ 8.0}}
& \hly{\textbf{67.2 $\pm$ 6.7}}
& \hly{\textbf{81.3 $\pm$ 1.2}}
& \hly{\textbf{71.6 $\pm$ 4.3}}
& \hly{\textbf{53.7 $\pm$ 2.9}}
& \hly{\textbf{77.4 $\pm$ 3.3}}
& \hly{\textbf{89.1 $\pm$ 0.4}}
& \hly{\textbf{36.6 $\pm$ 5.5}}
& \hly{\textbf{95.1 $\pm$ 1.1}}
& \hly{\textbf{79.0 $\pm$ 0.5}}
& {1.0} \\

\bottomrule
\end{tabular}
}
\vspace{-3mm}
\end{table*}

%% file: Tabs/t6.tex
\begin{table*}[t]
\centering
\caption{{Accuracy comparison of {\hbc} and 22 competing methods with node classification on 10 real-world hypergraphs}}
\label{tab_6}
\normalsize
% \vspace{-2mm}
\def\arraystretch{1.3}
\resizebox{\textwidth}{!}{
\begin{tabular}{cc||cccccccccc|c}
\toprule
&& \textbf{Citeseer} & \textbf{Cora} & \textbf{DBLP-A}
& \textbf{Cora-CA} & \textbf{IMDB} & \textbf{Pubmed}
& \textbf{DBLP-P} & \textbf{Aminer} & \textbf{MN-40}
& \textbf{20News} & \textbf{A.R.} \\
\midrule
\midrule

\multirow{19}{*}{\rotatebox{90}{\textbf{Euclidean}}}
& MLP
& 32.4 $\pm$ 8.1 & 29.4 $\pm$ 5.3&56.6 $\pm$ 5.4
& 36.0 $\pm$ 4.7 & 37.6 $\pm$ 3.3&62.8 $\pm$ 3.0
& 74.6 $\pm$ 0.8 & 22.7 $\pm$ 1.8&88.5 $\pm$ 1.5
& 73.3 $\pm$ 1.3 & 18.9 \\

& HGNN
&38.1 $\pm$ 10.7&45.1 $\pm$ 9.1&65.0 $\pm$ 5.7
&44.3 $\pm$ 8.0&41.5 $\pm$ 2.8&70.8 $\pm$ 3.8
&84.0 $\pm$ 0.9&29.7 $\pm$ 2.0&89.5 $\pm$ 1.3
&67.1 $\pm$ 2.0& 13.8 \\

& HyperGCN
&33.8 $\pm$ 7.3&39.1 $\pm$ 8.2&64.3 $\pm$ 6.3
&45.0 $\pm$ 9.5&40.6 $\pm$ 3.1&59.3 $\pm$ 15.2
&83.6 $\pm$ 1.6&26.8 $\pm$ 2.2&57.8 $\pm$ 11.6
&71.7 $\pm$ 2.1& 17.7 \\

& HNHN
&44.2 $\pm$ 7.9&52.5 $\pm$ 7.3&69.0 $\pm$ 4.5
&53.1 $\pm$ 7.4&42.5 $\pm$ 3.6&69.1 $\pm$ 3.4
&86.4 $\pm$ 0.8&32.1 $\pm$ 2.7&89.2 $\pm$ 1.4
&74.4 $\pm$ 1.1& 9.7 \\

& UniGCN
&39.8 $\pm$ 9.2&43.1 $\pm$ 6.6&59.4 $\pm$ 5.1
&46.3 $\pm$ 6.6&41.0 $\pm$ 3.0&67.6 $\pm$ 5.6
&82.2 $\pm$ 0.9&29.8 $\pm$ 1.8&75.7 $\pm$ 11.4
&67.4 $\pm$ 2.5& 16.1 \\

& UniGIN
&41.4 $\pm$ 10.3&45.5 $\pm$ 7.9&63.1 $\pm$ 5.5
&49.2 $\pm$ 6.9&41.7 $\pm$ 2.7&70.4 $\pm$ 4.9
&83.5 $\pm$ 0.8&30.8 $\pm$ 1.2&87.1 $\pm$ 2.3
&68.1 $\pm$ 2.4& 13.2 \\

& UniGCN2
&39.6 $\pm$ 10.7&43.9 $\pm$ 9.5&58.7 $\pm$ 7.9
&55.3 $\pm$ 5.3&41.6 $\pm$ 3.9&72.6 $\pm$ 3.8
&86.1 $\pm$ 0.6&32.3 $\pm$ 1.5&79.7 $\pm$ 6.0
&76.8 $\pm$ 1.4& 11.4 \\

& AllSet
&41.1 $\pm$ 9.8&49.0 $\pm$ 8.1&64.9 $\pm$ 4.6
&53.6 $\pm$ 8.2&41.7 $\pm$ 2.9&74.1 $\pm$ 5.0
&85.6 $\pm$ 0.7&29.8 $\pm$ 3.2&89.4 $\pm$ 1.3
&\hlb{\mbox{\underline{77.2 $\pm$ 0.9}}}& 9.4 \\

& ED-HNN
&32.9 $\pm$ 11.6&29.3 $\pm$ 3.9&58.3 $\pm$ 5.2
&36.3 $\pm$ 8.7&37.2 $\pm$ 2.6&61.9 $\pm$ 3.3
&85.1 $\pm$ 1.1&27.1 $\pm$ 2.8&73.0 $\pm$ 3.8
&75.7 $\pm$ 1.2& 18.0 \\

& PhenomNN
&42.2 $\pm$ 9.0&55.8 $\pm$ 8.0&70.3 $\pm$ 5.3
&56.2 $\pm$ 6.6&42.1 $\pm$ 2.6
&\hlb{\mbox{\underline{76.8 $\pm$ 3.4}}}
& O.O.M & O.O.M &94.0 $\pm$ 0.5& O.O.T & 7.0 \\

& GraphMAE2
&51.7 $\pm$ 12.2&61.0 $\pm$ 9.1&77.2 $\pm$ 2.4
&64.3 $\pm$ 6.3&45.6 $\pm$ 3.1&72.6 $\pm$ 4.4
&87.4 $\pm$ 0.4&34.7 $\pm$ 2.3&90.6 $\pm$ 1.2
&71.8 $\pm$ 0.4& 6.2 \\

& MaskGAE
&53.2 $\pm$ 10.8&55.8 $\pm$ 9.0&78.2 $\pm$ 2.7
&59.8 $\pm$ 4.1&45.0 $\pm$ 3.0&75.4 $\pm$ 3.2
&86.9 $\pm$ 0.5&33.8 $\pm$ 1.7&91.0 $\pm$ 0.8
& O.O.T & 5.6 \\

& GGD
&34.0 $\pm$ 5.8&37.2 $\pm$ 6.7&58.8 $\pm$ 6.5
&32.2 $\pm$ 6.3&37.6 $\pm$ 3.2&64.9 $\pm$ 4.4
&50.5 $\pm$ 6.4&31.5 $\pm$ 1.7&76.8 $\pm$ 2.0
&47.5 $\pm$ 7.9& 18.3 \\

& TriCL
&53.0 $\pm$ 9.9&61.5 $\pm$ 5.3&80.3 $\pm$ 3.5
&63.4 $\pm$ 3.9&47.5 $\pm$ 4.6&74.0 $\pm$ 5.2
&87.6 $\pm$ 0.5&34.6 $\pm$ 2.7&93.0 $\pm$ 0.8
&73.8 $\pm$ 1.8& 5.2 \\

& HyperGCL
&43.4 $\pm$ 8.4&54.8 $\pm$ 5.6&69.5 $\pm$ 3.7
&61.8 $\pm$ 3.0&48.4 $\pm$ 2.5&71.0 $\pm$ 2.2
& O.O.M &30.4 $\pm$ 1.8
&\hlb{\mbox{\underline{94.6 $\pm$ 1.2}}}
&74.0 $\pm$ 1.9& 8.0 \\

& HyperGRL
&35.1 $\pm$ 7.6&37.0 $\pm$ 5.3&41.1 $\pm$ 5.1
&41.8 $\pm$ 5.8&35.7 $\pm$ 4.2&50.2 $\pm$ 8.7
&78.7 $\pm$ 1.2&28.0 $\pm$ 2.8&89.4 $\pm$ 1.5
& O.O.T & 18.7 \\

& HypeBoy
& \hlb{\mbox{\underline{57.7 $\pm$ 10.8}}}
&61.9 $\pm$ 8.5
&\hlb{\mbox{\underline{81.2 $\pm$ 2.7}}}
&\hlb{\mbox{\underline{67.0 $\pm$ 3.7}}}
&48.3 $\pm$ 5.0&73.7 $\pm$ 4.1
&\hlb{\mbox{\underline{87.9 $\pm$ 0.5}}}
&34.6 $\pm$ 2.8&89.2 $\pm$ 2.2
&75.7 $\pm$ 2.1& 4.2 \\

& VilLain
&33.6 $\pm$ 9.4&47.1 $\pm$ 8.4&40.4 $\pm$ 3.9
&31.4 $\pm$ 4.1&39.7 $\pm$ 4.4&73.7 $\pm$ 3.3
&69.0 $\pm$ 0.9&19.9 $\pm$ 1.2&69.7 $\pm$ 2.3
&74.3 $\pm$ 3.0& 17.7 \\

& SE-HSSL
&57.0 $\pm$ 8.2
&\hlb{\mbox{\underline{62.5 $\pm$ 6.2}}}
&77.1 $\pm$ 2.7&63.9 $\pm$ 5.0
&\hlb{\mbox{\underline{48.9 $\pm$ 4.1}}}
&69.8 $\pm$ 2.8&83.6 $\pm$ 0.9
&\hlb{\mbox{\underline{34.8 $\pm$ 4.2}}}
&85.8 $\pm$ 5.2&75.8 $\pm$ 1.0& 6.9 \\

\midrule

\multirow{4}{*}{\rotatebox{90}{\textbf{Hyperbolic}}}
& HB-GCN
&36.8 $\pm$ 7.1&41.4 $\pm$ 5.7&64.3 $\pm$ 3.5
&52.7 $\pm$ 6.7&40.1 $\pm$ 1.8&71.9 $\pm$ 4.0
&85.7 $\pm$ 0.8&30.3 $\pm$ 1.1&86.0 $\pm$ 2.1
&74.6 $\pm$ 0.8& 13.2 \\

& HB-GCL
&57.3 $\pm$ 2.6&57.2 $\pm$ 3.7&48.7 $\pm$ 5.1
&63.6 $\pm$ 1.9&44.5 $\pm$ 2.4&65.7 $\pm$ 3.7
& O.O.M &25.7 $\pm$ 1.7&61.4 $\pm$ 4.7
&61.0 $\pm$ 3.9& 13.6 \\

& THINK
&42.5 $\pm$ 8.7&45.0 $\pm$ 8.5&55.6 $\pm$ 4.6
&46.9 $\pm$ 5.2&41.2 $\pm$ 4.2&74.0 $\pm$ 4.4
&85.0 $\pm$ 1.0&29.9 $\pm$ 2.6&86.8 $\pm$ 1.8
&76.7 $\pm$ 1.3& 12.2 \\

& \textbf{HyBridge}
& \hly{\textbf{60.7 $\pm$ 8.0}} & \hly{\textbf{67.2 $\pm$
6.7}} & \hly{\textbf{81.3 $\pm$ 1.2}}
& \hly{\textbf{71.6 $\pm$ 4.3}} & \hly{\textbf{53.7 $\pm$ 2.9}} & \hly{\textbf{77.4 $\pm$ 3.3}}
& \hly{\textbf{89.1 $\pm$ 0.4}} & \hly{\textbf{36.6 $\pm$ 5.5}} & \hly{\textbf{95.1 $\pm$ 1.1}}
& \hly{\textbf{79.0 $\pm$ 0.5}} & \textbf{1.0} \\
\midrule
\midrule
\multicolumn{2}{c||}{\textbf{Improvement}} & \textcolor{blue}{+5.20\%}   &\textcolor{blue}{+7.52\%}  &\textcolor{blue}{+0.12\%}  & \textcolor{blue}{+6.87\%} & \textcolor{blue}{+9.82\%} &\textcolor{blue}{+0.78\%}  &\textcolor{blue}{+1.37\%}  &\textcolor{blue}{+5.17\%}  & \textcolor{blue}{+0.53\%}&\textcolor{blue}{+2.33\%}  &  \\
\bottomrule
\end{tabular}
}
\vspace{-4mm}
\end{table*}

%% file: Tabs/t7.tex
\begin{table*}[t]
\centering
\caption{AUROC comparison of \hbc \ and 22 competing methods with hyperedge prediction on 10 real-world hypergraphs}
\label{tab_7}
\normalsize
% \vspace{-2mm}
\def\arraystretch{1.3}

\resizebox{\textwidth}{!}{%
\begin{tabular}{cc||cccccccccc|c}
\toprule[1.3pt]
&& \textbf{Citeseer}
& \textbf{Cora}
& \textbf{DBLP-A}
& \textbf{Cora-CA}
& \textbf{IMDB}
& \textbf{Pubmed}
& \textbf{DBLP-P}
& \textbf{Aminer}
& \textbf{MN-40}
& \textbf{20News}
& \textbf{A.R.} \\
\midrule[1.0pt]

% ================= Euclidean =================
\multirow{19}{*}{\rotatebox{90}{\textbf{Euclidean}}}

& MLP
& $65.6 \pm 2.0$
& $77.5 \pm 1.5$
& $74.2 \pm 1.5$
& $58.6 \pm 2.9$
& $42.0 \pm 2.2$
& $88.3 \pm 0.6$
& $90.2 \pm 0.5$
& $82.3 \pm 1.0$
& $94.9 \pm 8.4$
& $95.1 \pm 3.9$
& 12.0 \\

& HGNN
& $54.8 \pm 4.1$
& $57.6 \pm 6.0$
& $64.7 \pm 1.3$
& $65.0 \pm 4.6$
& \hly{\textbf{60.5 $\pm$ 1.9}}
& $65.9 \pm 2.1$
& $61.0 \pm 8.4$
& $69.6 \pm 1.2$
& $50.2 \pm 0.6$
& $49.4 \pm 8.2$
& 15.5 \\

& HyperGCN
& $54.9 \pm 2.1$
& $64.9 \pm 2.0$
& $66.5 \pm 1.2$
& $49.3 \pm 1.5$
& $48.4 \pm 1.9$
& $74.1 \pm 0.6$
& $53.9 \pm 2.1$
& $50.8 \pm 2.1$
& $60.9 \pm 3.0$
& $55.6 \pm 15.6$
& 17.3 \\

& HNHN
& $53.7 \pm 2.5$
& $59.0 \pm 1.8$
& $62.4 \pm 1.9$
& $51.6 \pm 1.4$
& $48.2 \pm 0.7$
& $65.7 \pm 2.4$
& $51.2 \pm 1.6$
& $48.1 \pm 1.3$
& $61.0 \pm 4.3$
& $56.0 \pm 9.7$
& 18.6 \\

& UniGCN
& $54.7 \pm 3.2$
& $51.6 \pm 2.0$
& $51.8 \pm 2.1$
& $60.0 \pm 5.9$
& $53.7 \pm 4.1$
& $52.1 \pm 3.2$
& $67.0 \pm 1.5$
& $69.0 \pm 1.3$
& $51.2 \pm 2.4$
& $50.0 \pm 0.0$
& 18.2 \\

& UniGIN
& $52.9 \pm 3.0$
& $63.1 \pm 3.7$
& $54.0 \pm 3.0$
& $58.7 \pm 8.2$
& $54.6 \pm 4.2$
& $51.5 \pm 3.6$
& $69.2 \pm 0.6$
& $69.1 \pm 1.4$
& $50.4 \pm 1.7$
& $50.0 \pm 0.0$
& 17.7 \\

& UniGCN2
& $60.2 \pm 2.1$
& $63.5 \pm 3.5$
& $61.6 \pm 1.7$
& $51.9 \pm 4.3$
& $57.5 \pm 7.0$
& $69.1 \pm 1.2$
& $60.2 \pm 2.5$
& $45.9 \pm 4.9$
& $51.7 \pm 1.9$
& $50.0 \pm 0.0$
& 16.6 \\

& AllSet
& $51.8 \pm 3.2$
& $54.5 \pm 3.4$
& $60.5 \pm 3.0$
& $51.2 \pm 3.2$
& $50.3 \pm 1.0$
& $52.3 \pm 1.6$
& $53.2 \pm 1.5$
& $51.0 \pm 1.8$
& $56.4 \pm 1.6$
& $50.1 \pm 0.4$
& 19.5 \\

& ED-HNN
& $58.0 \pm 3.8$
& $72.2 \pm 2.1$
& $61.9 \pm 6.4$
& $58.9 \pm 2.1$
& $49.4 \pm 1.2$
& $53.2 \pm 2.7$
& $51.2 \pm 1.6$
& $56.0 \pm 3.7$
& $52.7 \pm 1.6$
& $51.4 \pm 3.5$
& 17.2 \\

& PhenomNN
& $54.5 \pm 2.7$
& $56.3 \pm 1.4$
& $64.6 \pm 2.0$
& $50.3 \pm 1.9$
& $49.7 \pm 0.9$
& $64.0 \pm 4.6$
& O.O.M
& O.O.M
& $63.5 \pm 8.6$
& O.O.T
& 18.4 \\

& GraphMAE2
& $78.2 \pm 2.3$
& $82.0 \pm 1.4$
& $83.3 \pm 1.4$
& $74.7 \pm 2.7$
& $47.2 \pm 1.0$
& $93.6 \pm 2.6$
& $92.7 \pm 0.3$
& $78.1 \pm 0.8$
& \hlb{\mbox{$\underline{99.8 \pm 0.0}$}}
& $87.0 \pm 3.4$
& 8.7 \\

& MaskGAE
& $86.7 \pm 1.8$
& $86.7 \pm 1.6$
& $87.3 \pm 1.3$
& $76.5 \pm 2.1$
& $54.4 \pm 1.7$
& \hlb{\mbox{$\underline{95.4 \pm 0.3}$}}
& $94.8 \pm 0.4$
& $87.6 \pm 0.9$
& \hly{\textbf{99.9 $\pm$ 0.0}}
& O.O.T
& 4.6 \\

& GGD
& $72.2 \pm 5.0$
& $71.9 \pm 3.1$
& $81.4 \pm 1.9$
& $73.2 \pm 4.0$
& $53.1 \pm 1.8$
& $87.2 \pm 0.7$
& $91.6 \pm 1.0$
& $84.9 \pm 2.1$
& $99.5 \pm 0.1$
& $87.9 \pm 3.1$
& 9.7 \\

& TriCL
& $90.5 \pm 1.2$
& $90.7 \pm 1.3$
& \hlb{\mbox{$\underline{87.9 \pm 1.4}$}}
& $87.8 \pm 1.5$
& $58.9 \pm 2.1$
& $91.9 \pm 0.5$
& $94.8 \pm 0.2$
& \hlb{\mbox{$\underline{90.4 \pm 0.6}$}}
& $99.6 \pm 0.1$
& \hlb{\mbox{$\underline{98.2 \pm 3.0}$}}
& 3.3 \\

& HyperGCL
& $73.9 \pm 2.6$
& $85.4 \pm 1.5$
& $83.5 \pm 1.0$
& $81.1 \pm 1.9$
& $53.8 \pm 2.4$
& $89.6 \pm 0.5$
& $83.6 \pm 0.6$
& $82.1 \pm 7.6$
& $99.4 \pm 0.1$
& $76.3 \pm 6.3$
& 8.7 \\

& HyperGRL
& $83.2 \pm 8.0$
& $84.0 \pm 1.7$
& $80.8 \pm 1.4$
& $80.0 \pm 3.3$
& $54.7 \pm 2.8$
& $78.3 \pm 1.8$
& $88.2 \pm 0.3$
& $81.5 \pm 0.7$
& $98.8 \pm 0.4$
& O.O.T
& 9.1 \\

& VilLain
& $58.3 \pm 2.7$
& $68.5 \pm 1.9$
& $57.3 \pm 2.0$
& $64.4 \pm 3.2$
& $43.9 \pm 2.4$
& $83.7 \pm 0.7$
& $56.3 \pm 0.7$
& $62.7 \pm 1.3$
& $79.8 \pm 0.7$
& $59.2 \pm 7.8$
& 15.8 \\

& SE-HSSL
& \hlb{\mbox{$\underline{90.5 \pm 1.1}$}}
& \hlb{\mbox{$\underline{91.9 \pm 1.3}$}}
& $87.4 \pm 1.2$
& $85.6 \pm 2.1$
& $55.6 \pm 2.0$
& $94.5 \pm 0.7$
& $94.0 \pm 0.4$
& $87.5 \pm 0.6$
& $99.8 \pm 0.1$
& $95.2 \pm 5.8$
& 4.1 \\

& HypeBoy
& $86.4 \pm 2.1$
& $91.1 \pm 1.2$
& $87.0 \pm 1.2$
& $87.3 \pm 1.6$
& $59.4 \pm 1.9$
& $92.1 \pm 0.5$
& \hlb{\mbox{$\underline{95.2 \pm 0.2}$}}
& $89.7 \pm 0.6$
& $99.7 \pm 0.0$
& $97.2 \pm 3.2$
& 4.1 \\

\midrule[1.2pt]

% ================= Hyperbolic =================
\multirow{4}{*}{\rotatebox{90}{\textbf{Hyperbolic}}}

& HB-GCN
& $66.9 \pm 3.3$
& $76.0 \pm 4.1$
& $66.9 \pm 1.8$
& $74.7 \pm 3.9$
& $54.0 \pm 3.1$
& $91.8 \pm 0.5$
& $90.5 \pm 0.8$
& $69.7 \pm 3.8$
& $97.3 \pm 1.0$
& $84.0 \pm 7.1$
& 10.2 \\

& HB-GCL
& $89.0 \pm 1.8$
& $85.4 \pm 1.2$
& $81.8 \pm 1.1$
& \hlb{\mbox{$\underline{87.8 \pm 1.4}$}}
& $53.6 \pm 3.2$
& $95.3 \pm 0.5$
& $81.8 \pm 1.1$
& $88.0 \pm 0.9$
& $99.6 \pm 0.1$
& $97.5 \pm 2.9$
& 6.2 \\

& THINK
& $62.5 \pm 6.7$
& $68.4 \pm 4.0$
& $65.7 \pm 3.1$
& $63.8 \pm 6.9$
& $49.4 \pm 3.0$
& $74.3 \pm 3.7$
& $84.4 \pm 2.3$
& $64.4 \pm 8.2$
& $98.7 \pm 2.3$
& $65.2 \pm 10.6$
& 13.6 \\

& \textbf{{\hbc}}
& \hly{\textbf{91.7 $\pm$ 1.1}}
& \hly{\textbf{93.0 $\pm$ 1.2}}
& \hly{\textbf{89.0 $\pm$ 0.9}}
& \hly{\textbf{89.9 $\pm$ 1.4}}
& \hlb{\mbox{$\underline{60.5 \pm 2.7}$}}
& \hly{\textbf{95.4 $\pm$ 0.2}}
& \hly{\textbf{95.6 $\pm$ 0.3}}
& \hly{\textbf{90.4 $\pm$ 0.5}}
& \hly{\textbf{99.9 $\pm$ 0.0}}
& \hly{\textbf{98.8 $\pm$ 1.6}}
& \textbf{1.1} \\
% \midrule
% \midrule
% \multicolumn{2}{c||}{\textbf{Improvement}} & \textcolor{blue}{+5.20\%}   &\textcolor{blue}{+7.52\%}  &\textcolor{blue}{+0.12\%}  & \textcolor{blue}{+6.87\%} & \textcolor{blue}{+9.82\%} &\textcolor{blue}{+0.78\%}  &\textcolor{blue}{+1.37\%}  &\textcolor{blue}{+5.17\%}  & \textcolor{blue}{+0.53\%}&\textcolor{blue}{+2.33\%}  &  \\
\bottomrule[1.2pt]
\end{tabular}%
}
\vspace{-3mm}
\end{table*}

%% file: Tabs/t8.tex
\begin{table*}[t]
\centering
\caption{Node classification accuracy of \hbc \ according to the label distributions used for bridge score estimation}
\label{tab_8}
\normalsize
\vspace{-2mm}
\def\arraystretch{1.3}

\resizebox{\textwidth}{!}{%
\begin{tabular}{c||cccccccccc|c}
\toprule[1.2pt]
\textbf{Label distribution}
& \textbf{Citeseer}
& \textbf{Cora}
& \textbf{DBLP-A}
& \textbf{Cora-CA}
& \textbf{IMDB}
& \textbf{Pubmed}
& \textbf{DBLP-P}
& \textbf{Aminer}
& \textbf{MN-40}
& \textbf{20News}
& \textbf{A.R.} \\
\midrule[1.0pt]

\textbf{neighborhood}
&  \hlb{\mbox{$\underline{59.3 \pm 7.4}$}}
&  \hlb{\mbox{$\underline{65.0 \pm 8.0}$}}
&  \hlb{\mbox{$\underline{77.0 \pm 2.3}$}}
&  \hlb{\mbox{$\underline{69.5 \pm 3.9}$}}
&  \hlb{\mbox{$\underline{53.3 \pm 3.2}$}}
&  \hlb{\mbox{$\underline{76.6 \pm 3.8}$}}
&  \hlb{\mbox{$\underline{88.1 \pm 0.5}$}}
&  \hlb{\mbox{$\underline{36.2 \pm 5.8}$}}
&  \hlb{\mbox{$\underline{95.1 \pm 1.2}$}}
&  \hlb{\mbox{$\underline{78.1 \pm 0.7}$}}
& 2.0 \\

\textbf{external}
& \hly{\textbf{60.7 $\pm$ 8.0}}
& \hly{\textbf{67.2 $\pm$ 6.7}}
& \hly{\textbf{81.3 $\pm$ 1.2}}
& \hly{\textbf{71.6 $\pm$ 4.3}}
& \hly{\textbf{53.7 $\pm$ 2.9}}
& \hly{\textbf{77.4 $\pm$ 3.3}}
& \hly{\textbf{89.1 $\pm$ 0.4}}
& \hly{\textbf{36.6 $\pm$ 5.5}}
& \hly{\textbf{95.1 $\pm$ 1.1}}
& \hly{\textbf{79.0 $\pm$ 0.5}}
& 1.0 \\

\bottomrule[1.2pt]
\end{tabular}%
}
\vspace{-2mm}
\end{table*}

%% file: Figs/f6.tex
\begin{figure}[t]
    \centering
    \scriptsize

    \begin{tikzpicture}
    \begin{groupplot}[
        group style={
            group size=2 by 1,
            horizontal sep=0.75cm,
            vertical sep=1.0cm
        },
        width=5.3cm, height=3cm,
        xmin=0, xmax=9,
        xtick={0,1,2,3,4,5,6,7,8,9},
        grid=both,
        grid style={gray!20},
        tick align=outside,
        tick pos=left,
        label style={font=\scriptsize},
        tick label style={font=\scriptsize},
        xlabel style={font=\scriptsize},
        every axis plot/.append style={
            line width=0.85pt,
            mark size=0.45pt
        },
        legend columns=3,
        legend style={
            at={(1.08,1.20)},
            anchor=south,
            draw=none,
            font=\scriptsize,
            /tikz/every even column/.append style={
                column sep=0.25cm
            }
        },
        legend cell align=left,
        clip=false
    ]

    % =============== Label entropy ===============
    \nextgroupplot[
        ymin=0, ymax=0.9,
        xlabel={(a) Label entropy}
    ]

    % Citeseer
    \addplot+[blue, mark=*] coordinates {
        (0,0.1833408) (1,0.16566449)
        (2,0.16505721) (3,0.18842048)
        (4,0.16367883) (5,0.22958863)
        (6,0.21937974) (7,0.18025486)
        (8,0.16149122) (9,0.12978306)
    };
    \addlegendentry{\texttt{Citeseer}}

    % Cora
    \addplot+[red, mark=square*] coordinates {
        (0,0.14553513) (1,0.21838602)
        (2,0.28348501) (3,0.21317589)
        (4,0.22093823) (5,0.16144786)
        (6,0.16416463) (7,0.16519781)
        (8,0.08615759) (9,0.05199761)
    };
    \addlegendentry{\texttt{Cora}}

    % Cora-CA
    \addplot+[green!60!black, mark=triangle*] coordinates {
        (0,0.20853377) (1,0.28235286)
        (2,0.28536615) (3,0.2486788)
        (4,0.24074792) (5,0.20612819)
        (6,0.1854984) (7,0.17387725)
        (8,0.12862271) (9,0.05986077)
    };
    \addlegendentry{\texttt{Cora-CA}}

    % DBLP-A
    \addplot+[orange!80!black, mark=diamond*] coordinates {
        (0,0.44895162) (1,0.36149863)
        (2,0.27677033) (3,0.13477451)
        (4,0.09517893) (5,0.05254201)
        (6,0.06666395) (7,0.03225298)
        (8,0.01283373) (9,0.00547605)
    };
    \addlegendentry{\texttt{DBLP-A}}

    % IMDB
    \addplot+[purple, mark=pentagon*] coordinates {
        (0,0.75954111) (1,0.81358057)
        (2,0.78936108) (3,0.7886032)
        (4,0.77040973) (5,0.74702171)
        (6,0.71305279) (7,0.67154623)
        (8,0.67116695) (9,0.54945921)
    };
    \addlegendentry{\texttt{IMDB}}

    % Pubmed
    \addplot+[cyan!70!black, mark=asterisk] coordinates {
        (0,0.56458746) (1,0.50794662)
        (2,0.40755096) (3,0.37651167)
        (4,0.37195514) (5,0.40954815)
        (6,0.37825042) (7,0.41243105)
        (8,0.42878733) (9,0.29333473)
    };
    \addlegendentry{\texttt{Pubmed}}

    % ================= Homophily =================
    \nextgroupplot[
        ymin=0.4, ymax=1.0,
        xlabel={(b) Homophily}
    ]

    % Citeseer
    \addplot+[blue, mark=*] coordinates {
        (0,0.52038314) (1,0.62092747)
        (2,0.64708661) (3,0.60419404)
        (4,0.69868187) (5,0.66421977)
        (6,0.69395082) (7,0.73326274)
        (8,0.75687768) (9,0.7843028)
    };

    % Cora
    \addplot+[red, mark=square*] coordinates {
        (0,0.74436505) (1,0.70560926)
        (2,0.65345925) (3,0.68363507)
        (4,0.71837381) (5,0.78380117)
        (6,0.76498725) (7,0.74653314)
        (8,0.82440793) (9,0.9219031)
    };

    % Cora-CA
    \addplot+[green!60!black, mark=triangle*] coordinates {
        (0,0.66642369) (1,0.63753333)
        (2,0.69060224) (3,0.69597544)
        (4,0.68044597) (5,0.75479435)
        (6,0.75953447) (7,0.76313596)
        (8,0.82610302) (9,0.92464663)
    };

    % DBLP-A
    \addplot+[orange!80!black, mark=diamond*] coordinates {
        (0,0.51394231) (1,0.62518254)
        (2,0.68401693) (3,0.80509074)
        (4,0.80786756) (5,0.85384648)
        (6,0.8587608) (7,0.91692407)
        (8,0.93082368) (9,0.95688546)
    };

    % IMDB
    \addplot+[purple, mark=pentagon*] coordinates {
        (0,0.42894946) (1,0.43431477)
        (2,0.44695456) (3,0.43381484)
        (4,0.45087326) (5,0.46470233)
        (6,0.46647006) (7,0.49587256)
        (8,0.49858366) (9,0.55063471)
    };

    % Pubmed
    \addplot+[cyan!70!black, mark=asterisk] coordinates {
        (0,0.60343474) (1,0.67498424)
        (2,0.75402057) (3,0.76362722)
        (4,0.78123223) (5,0.74435532)
        (6,0.75486593) (7,0.74132036)
        (8,0.73532512) (9,0.79656905)
    };

    \end{groupplot}
    \end{tikzpicture}

    \vspace{-2mm}
    \caption{
        Label entropy and homophily across hyperbolic radius bins after applying bridge loss.
    }
    \label{f6}
\end{figure}
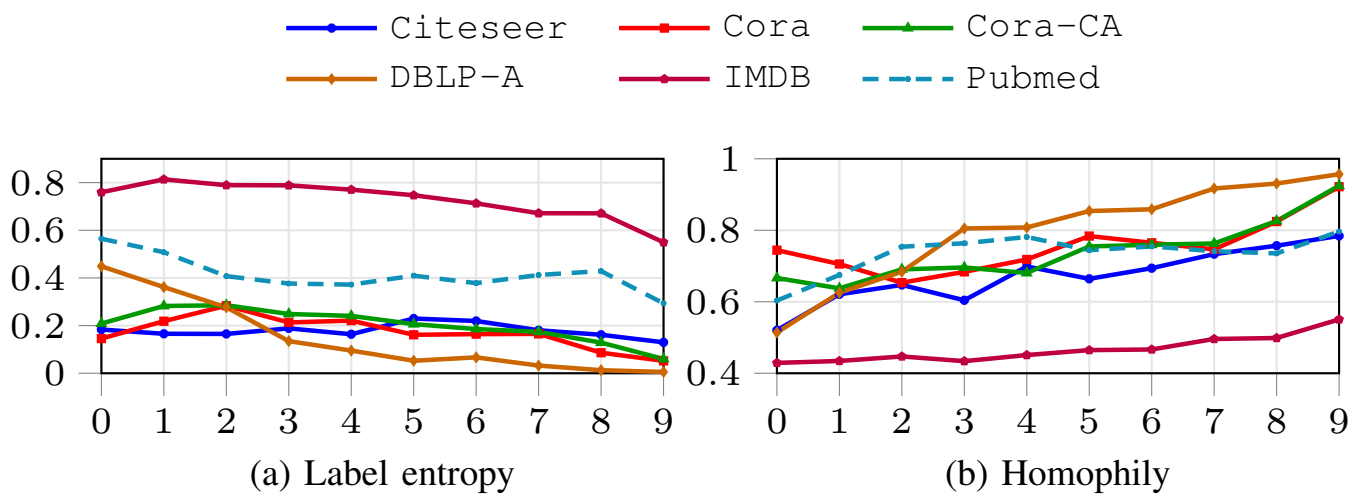

%% file: Figs/f8.tex
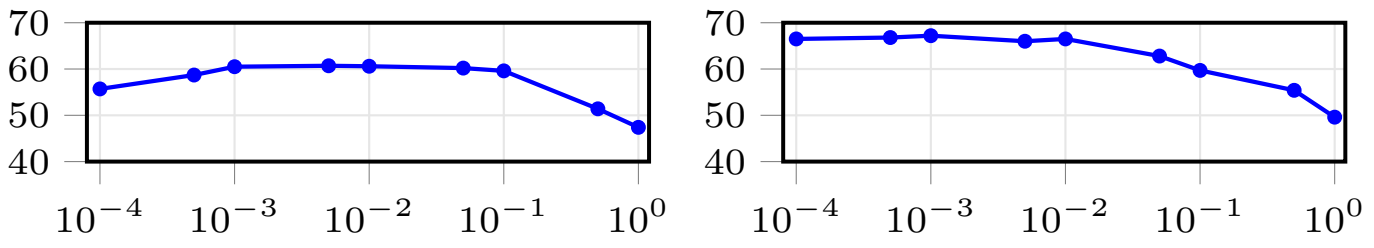
\begin{figure}[t]
\centering
\scriptsize

% ================= Citeseer =================
\begin{subfigure}[t]{0.48\linewidth}
\centering
\begin{tikzpicture}
\begin{axis}[
    width=5.3cm, height=2.5cm,
    xmode=log,
    log basis x=10,
    xmin=0.00008, xmax=1.2,
    ymin=40, ymax=70,
    xtick={0.0001,0.001,0.01,0.1,1},
    xticklabels={
        $10^{-4}$,
        $10^{-3}$,
        $10^{-2}$,
        $10^{-1}$,
        $10^{0}$
    },
    ytick={40,50,60,70},
    grid=both,
    grid style={gray!20},
    tick align=outside,
    tick pos=left,
    line width=0.75pt,
    label style={font=\scriptsize},
    tick label style={font=\scriptsize}
]

\addplot[
    blue,
    thick,
    mark=*,
    mark size=1pt
] coordinates {
    (0.0001,55.7)
    (0.0005,58.7)
    (0.001,60.5)
    (0.005,60.7)
    (0.01,60.6)
    (0.05,60.2)
    (0.1,59.6)
    (0.5,51.4)
    (1,47.4)
};

\end{axis}
\end{tikzpicture}
% \vspace{-3mm}
\caption{\footnotesize{Citeseer}}
\end{subfigure}
\hfill
% ================= Cora =================
\begin{subfigure}[t]{0.48\linewidth}
\centering
\begin{tikzpicture}
\begin{axis}[
    width=5.3cm, height=2.5cm,
    xmode=log,
    log basis x=10,
    xmin=0.00008, xmax=1.2,
    ymin=40, ymax=70,
    xtick={0.0001,0.001,0.01,0.1,1},
    xticklabels={
        $10^{-4}$,
        $10^{-3}$,
        $10^{-2}$,
        $10^{-1}$,
        $10^{0}$
    },
    ytick={40,50,60,70},
    grid=both,
    grid style={gray!20},
    tick align=outside,
    tick pos=left,
    line width=0.75pt,
    label style={font=\scriptsize},
    tick label style={font=\scriptsize}
]

\addplot[
    blue,
    thick,
    mark=*,
    mark size=1pt
] coordinates {
    (0.0001,66.5)
    (0.0005,66.8)
    (0.001,67.2)
    (0.005,66.0)
    (0.01,66.5)
    (0.05,62.8)
    (0.1,59.7)
    (0.5,55.4)
    (1,49.6)
};

\end{axis}
\end{tikzpicture}
% \vspace{-3mm}
\caption{\footnotesize{Cora}}
\end{subfigure}

\vspace{-2mm}
\caption{
Node classification accuracy of {\hbc} under different bridge loss weights.
}
\label{f8}
% \vspace{-1mm}
\end{figure}

%% file: Figs/f10.tex
\begin{figure}[t]
\centering
\scriptsize

% ================= Citeseer =================
\begin{subfigure}[t]{0.48\linewidth}
\centering
\begin{tikzpicture}
\begin{axis}[
    width=5.3cm, height=2.5cm,
    xmin=1, xmax=6,
    enlarge x limits=false,
    xtick={1,2,3,4,5,6},
    xticklabels={$2^{12}$,$2^{13}$,$2^{14}$,$2^{15}$,$2^{16}$,$2^{17}$},
    ymin=56, ymax=64,
    ytick={56,58,60,62,64},
    grid=both,
    grid style={gray!20},
    tick align=outside,
    tick pos=left,
    label style={font=\scriptsize},
    tick label style={font=\scriptsize}
]

\addplot[
    purple,
    thick,
    mark=*,
    mark size=1pt
] coordinates {
    (1,60.5)
    (2,60.5)
    (3,60.7)
    (4,60.7)
    (5,60.5)
    (6,60.5)
};

\end{axis}
\end{tikzpicture}
\vspace{-3mm}
\caption{\footnotesize{Citeseer}}
\end{subfigure}
\hfill
% ================= Cora =================
\begin{subfigure}[t]{0.48\linewidth}
\centering
\begin{tikzpicture}
\begin{axis}[
    width=5.3cm, height=2.5cm,
    xmin=1, xmax=6,
    enlarge x limits=false,
    xtick={1,2,3,4,5,6},
    xticklabels={$2^{12}$,$2^{13}$,$2^{14}$,$2^{15}$,$2^{16}$,$2^{17}$},
    ymin=61, ymax=69,
    ytick={61,63,65,67,69},
    grid=both,
    grid style={gray!20},
    tick align=outside,
    tick pos=left,
    label style={font=\scriptsize},
    tick label style={font=\scriptsize}
]

\addplot[
    purple,
    thick,
    mark=*,
    mark size=1pt
] coordinates {
    (1,66.2)
    (2,67.3)
    (3,67.1)
    (4,66.8)
    (5,66.8)
    (6,66.5)
};

\end{axis}
\end{tikzpicture}
\vspace{-3mm}
\caption{\footnotesize{Cora}}
\end{subfigure}

\vspace{-2mm}
\caption{
Node classification accuracy of {\hbc} under different numbers of sampled node pairs.
}
\label{f10}
% \vspace{-1mm}
\end{figure}
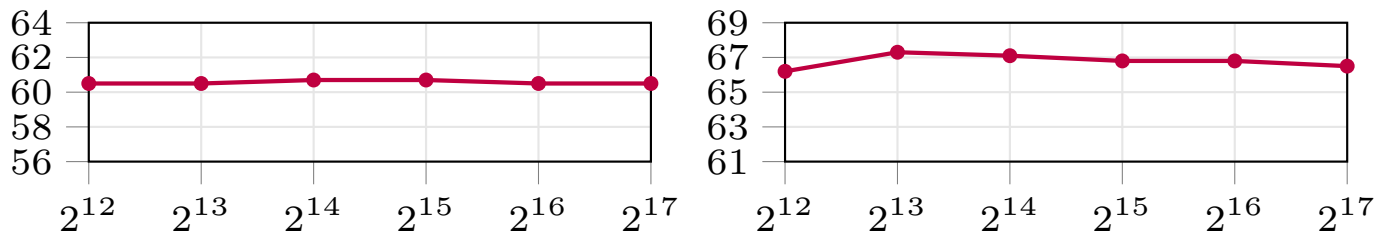

%% file: Figs/f9.tex
\begin{figure}[t]
\centering
\footnotesize
\begin{subfigure}[t]{0.19\textwidth}
    \centering
    \includegraphics[width=\linewidth]{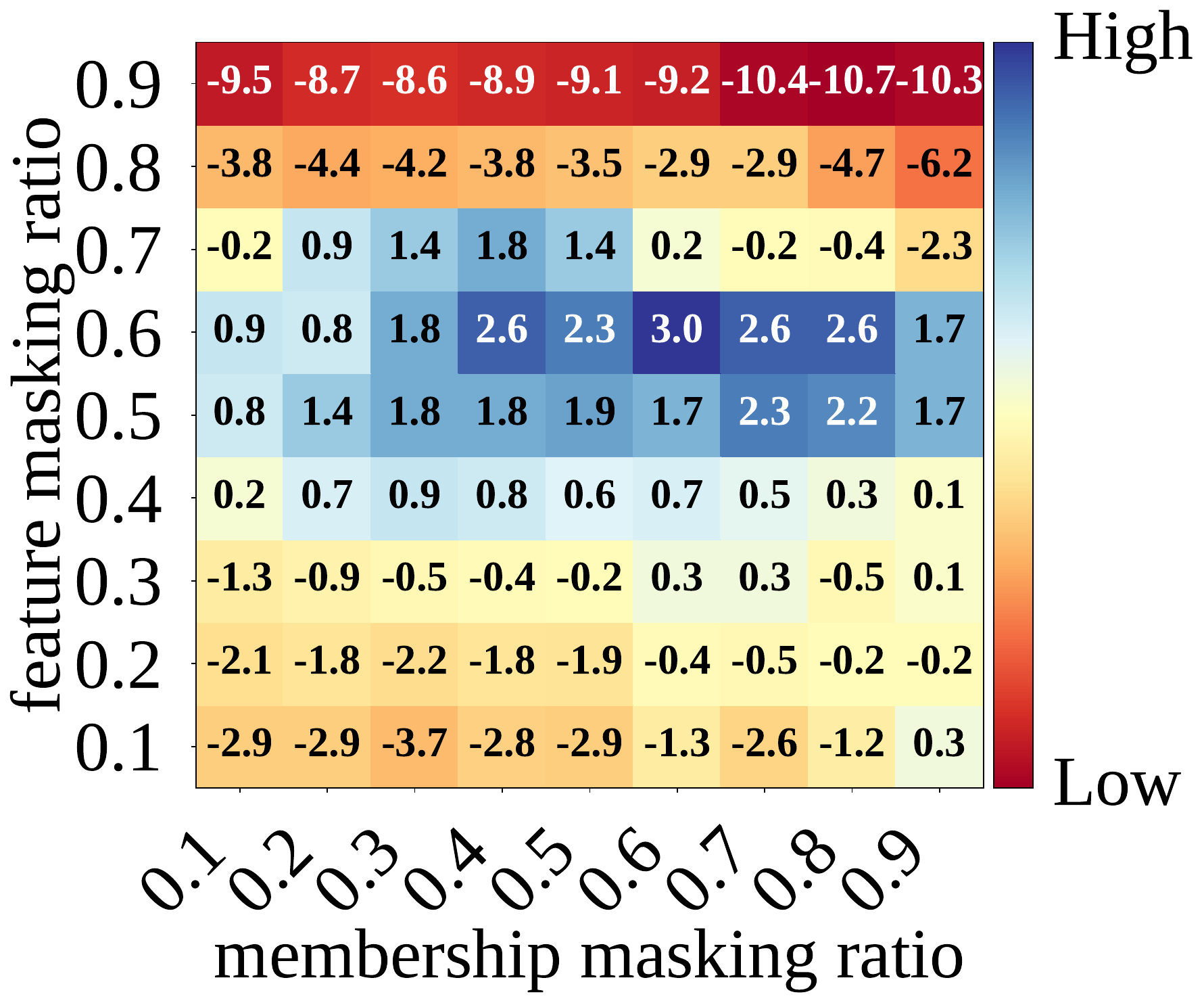}
            \vspace{-3mm}
    \caption{\normalsize{Citeseer}}
    \label{f8c}
\end{subfigure}
% \hfill
\hspace{3mm}
\begin{subfigure}[t]{0.19\textwidth}
    \centering
    \includegraphics[width=\linewidth]{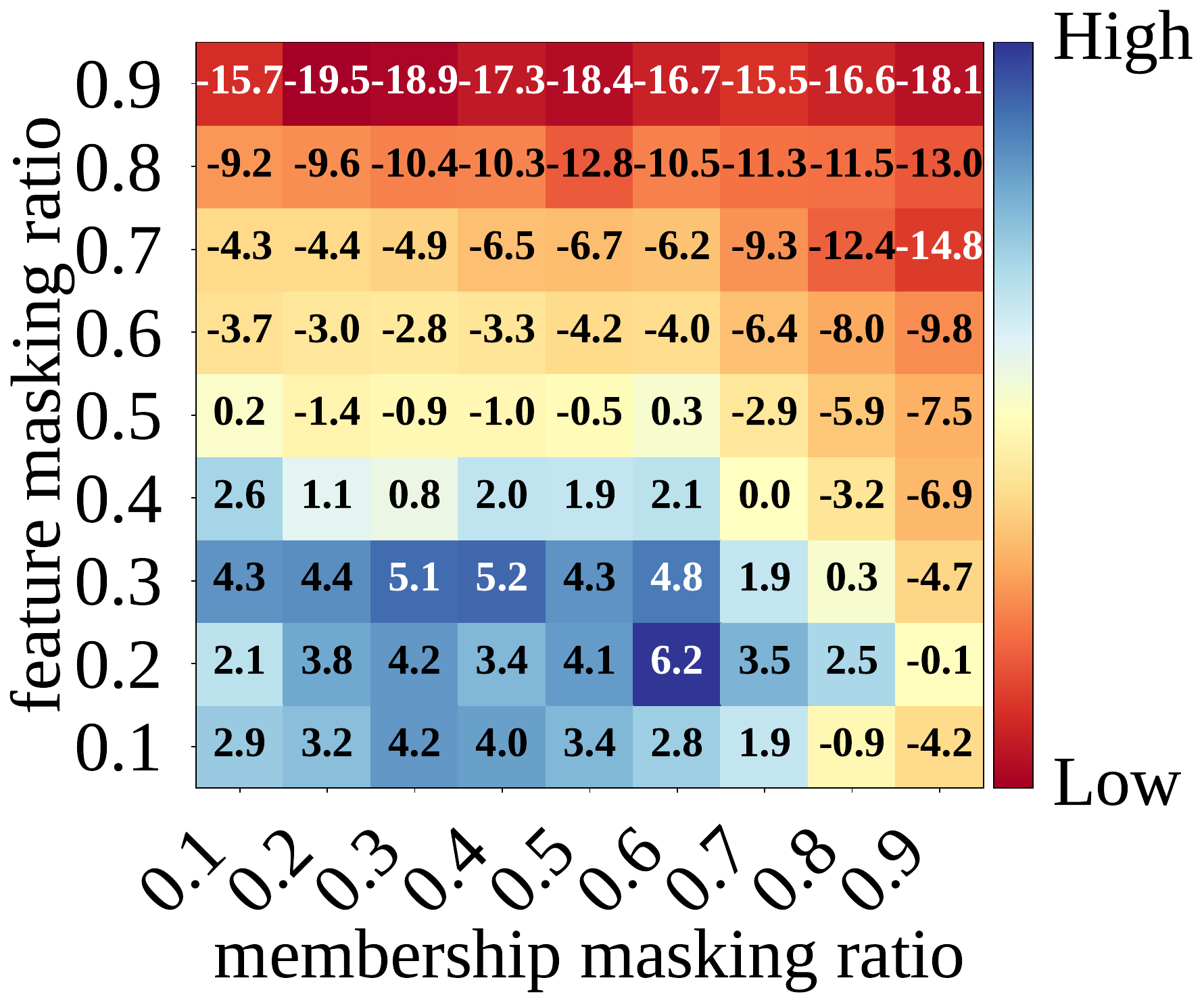}
        \vspace{-3mm}
    \caption{\normalsize{Cora}}
    \label{f8d}
\end{subfigure}
\vspace{-1mm}
\caption{Node classification accuracy of {\hbc} under different masking ratio.}
\label{f9}
\vspace{-4mm}
\end{figure}

%% file: Figs/f12.tex
\begin{figure}[t]
\centering
\scriptsize

\begin{tikzpicture}
\begin{axis}[
    ybar stacked,
    bar width=22pt,
    width=\columnwidth,
    height=3.0cm,
    ymin=0, ymax=2.8,
    ytick={0,1,2,3},
    ylabel={\shortstack{Processing\\time(sec)}},
    xmin=0.5, xmax=4.5,
    xtick={1,2,3,4},
    xticklabels={
        Citeseer,
        Cora,
        DBLP-P,
        AMiner
    },
    xticklabel style={
        rotate=35,
        anchor=east,
        font=\scriptsize
    },
    label style={font=\scriptsize},
    tick label style={font=\scriptsize},
    ymajorgrids=true,
    grid style={gray!30,dashed},
    tick align=outside,
    legend style={
        at={(0.98,0.98)},
        anchor=north east,
        font=\scriptsize,
        draw=gray!40,
        fill=white
    },
    legend cell align=left,
    enlarge x limits=false,
    clip=false
]

% ===== METIS =====
\addplot[
    fill=blue!75,
    draw=none
] coordinates {
    (1,0.005863)
    (2,0.007303)
    (3,0.330398)
    (4,0.096171)
};
\addlegendentry{METIS}

% ===== Bridge Score =====
\addplot[
    fill=orange!90,
    draw=none
] coordinates {
    (1,0.189292)
    (2,0.539133)
    (3,1.807823)
    (4,0.482840)
};
\addlegendentry{Bridge score}

% ===== Total Processing Time =====
% \node[font=\scriptsize,anchor=south]
%     at (axis cs:1,0.195154) {0.20};

% \node[font=\scriptsize,anchor=south]
%     at (axis cs:2,0.546436) {0.55};

% \node[font=\scriptsize,anchor=south]
%     at (axis cs:3,2.398221) {2.40};

% \node[font=\scriptsize,anchor=south]
%     at (axis cs:4,0.579011) {0.58};

\end{axis}
\end{tikzpicture}

\vspace{-2mm}
\caption{
Preprocessing time of METIS and bridge score
estimation.}
\label{f12}
% \vspace{-3mm}
\end{figure}
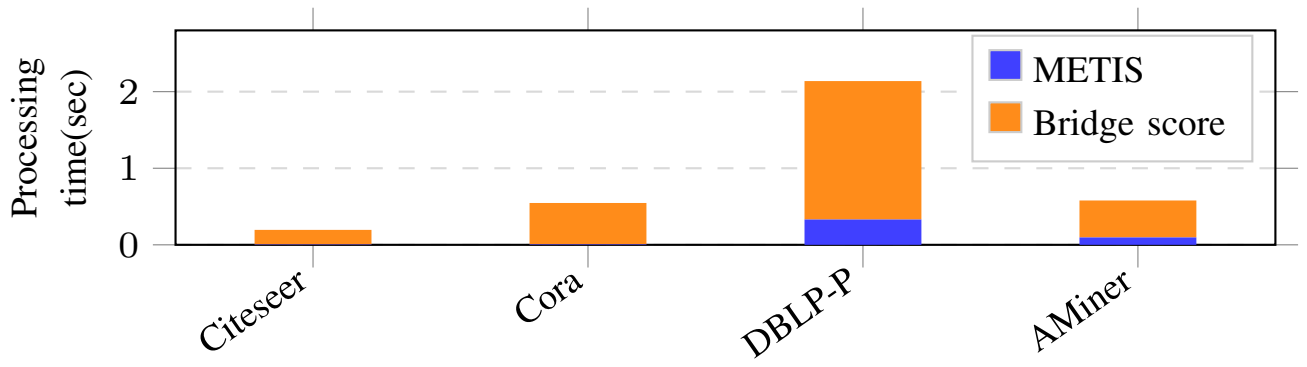

%% file: Figs/f13.tex
% Preamble
% \usepackage{graphicx}
% \usepackage{tikz}
% \usepackage{pgfplots}
% \usetikzlibrary{plotmarks}
% \pgfplotsset{compat=1.18}

% Filled five-pointed star: ★
\pgfdeclareplotmark{filledstar}{%
    \pgfmathsetlengthmacro{\innerstar}{0.42\pgfplotmarksize}
    \pgfpathmoveto{\pgfqpointpolar{90}{\pgfplotmarksize}}
    \pgfpathlineto{\pgfqpointpolar{54}{\innerstar}}
    \pgfpathlineto{\pgfqpointpolar{18}{\pgfplotmarksize}}
    \pgfpathlineto{\pgfqpointpolar{-18}{\innerstar}}
    \pgfpathlineto{\pgfqpointpolar{-54}{\pgfplotmarksize}}
    \pgfpathlineto{\pgfqpointpolar{-90}{\innerstar}}
    \pgfpathlineto{\pgfqpointpolar{-126}{\pgfplotmarksize}}
    \pgfpathlineto{\pgfqpointpolar{-162}{\innerstar}}
    \pgfpathlineto{\pgfqpointpolar{-198}{\pgfplotmarksize}}
    \pgfpathlineto{\pgfqpointpolar{-234}{\innerstar}}
    \pgfpathclose
    \pgfusepathqfillstroke
}

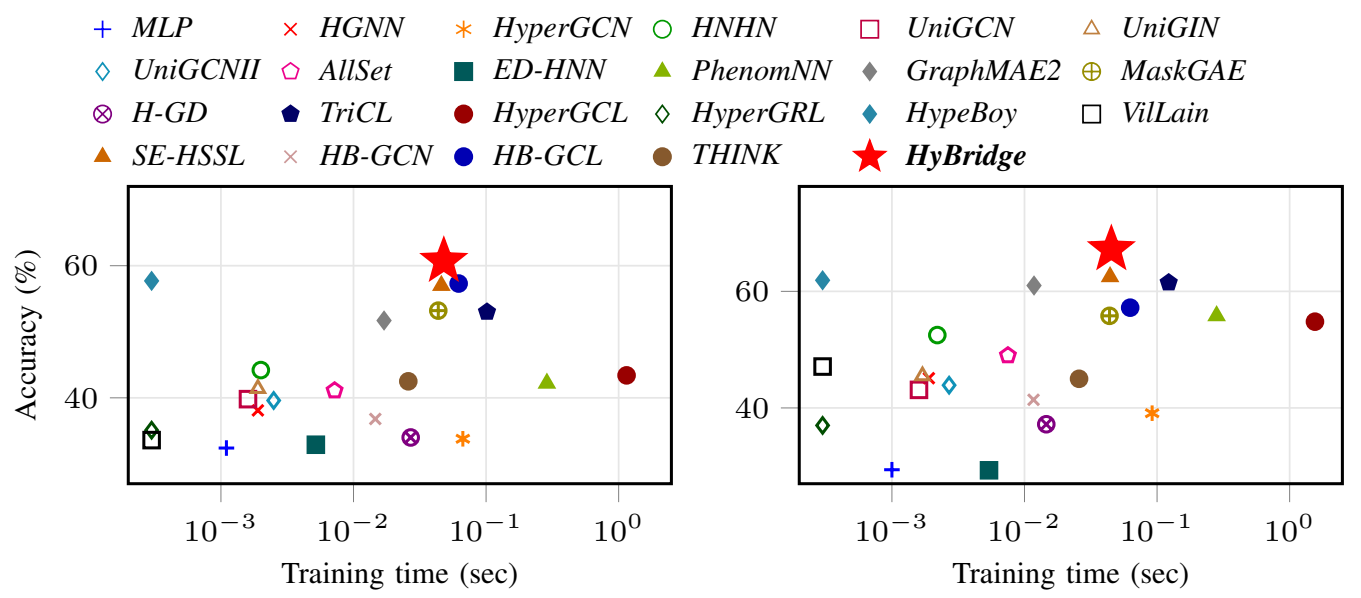
\begin{figure}[t]
\centering
\scriptsize

\newcommand{\timepoint}[4]{%
    \addplot[
        only marks,
        color=#1,
        mark=#2,
        mark size=1.8pt
    ] coordinates {(#3,#4)};
}

\resizebox{\columnwidth}{!}{%
\begin{tabular}{@{}cc@{}}

% ================= Legend =================
\multicolumn{2}{c}{
\begin{tikzpicture}
\begin{customlegend}[
    legend columns=6,
    legend style={
        align=center,
        draw=none,
        font=\scriptsize,
        column sep=0.10cm,
        /tikz/every even column/.append style={
            column sep=0.10cm
        }
    },
    legend cell align=left,
    legend entries={
        \textit{MLP},
        \textit{HGNN},
        \textit{HyperGCN},
        \textit{HNHN},
        \textit{UniGCN},
        \textit{UniGIN},
        \textit{UniGCNII},
        \textit{AllSet},
        \textit{ED-HNN},
        \textit{PhenomNN},
        \textit{GraphMAE2},
        \textit{MaskGAE},
        \textit{H-GD},
        \textit{TriCL},
        \textit{HyperGCL},
        \textit{HyperGRL},
        \textit{HypeBoy},
        \textit{VilLain},
        \textit{SE-HSSL},
        \textit{HB-GCN},
        \textit{HB-GCL},
        \textit{THINK},
        \textbf{\textit{HyBridge}}
    }
]
    \addlegendimage{only marks,color=blue,mark=+,mark size=2}
    \addlegendimage{only marks,color=red,mark=x,mark size=2}
    \addlegendimage{only marks,color=orange,mark=asterisk,mark size=2}
    \addlegendimage{only marks,color=green!60!black,mark=o,mark size=2}
    \addlegendimage{only marks,color=purple,mark=square,mark size=2}
    \addlegendimage{only marks,color=brown,mark=triangle,mark size=2}

    \addlegendimage{only marks,color=cyan!70!black,mark=diamond,mark size=2}
    \addlegendimage{only marks,color=magenta,mark=pentagon,mark size=2}
    \addlegendimage{only marks,color=teal!70!black,mark=square*,mark size=2}
    \addlegendimage{only marks,color=lime!70!black,mark=triangle*,mark size=2}
    \addlegendimage{only marks,color=gray,mark=diamond*,mark size=2}
    \addlegendimage{only marks,color=olive,mark=oplus,mark size=2}

    \addlegendimage{only marks,color=violet,mark=otimes,mark size=2}
    \addlegendimage{only marks,color=blue!40!black,mark=pentagon*,mark size=2}
    \addlegendimage{only marks,color=red!60!black,mark=*,mark size=2}
    \addlegendimage{only marks,color=green!30!black,mark=diamond,mark size=2}
    \addlegendimage{only marks,color=cyan!60!black,mark=diamond*,mark size=2}
    \addlegendimage{only marks,color=black,mark=square,mark size=2}

    \addlegendimage{only marks,color=orange!80!black,mark=triangle*,mark size=2}
    \addlegendimage{only marks,color=pink!80!black,mark=x,mark size=2}
    \addlegendimage{only marks,color=blue!70!black,mark=oplus*,mark size=2}
    \addlegendimage{only marks,color=brown!70!black,mark=otimes*,mark size=2}

    % HyBridge: filled red five-point star
    \addlegendimage{
        only marks,
        color=red,
        mark=filledstar,
        mark size=3.8pt,
        mark options={
            fill=red,
            draw=red
        }
    }
\end{customlegend}
\end{tikzpicture}
}
\\[-1mm]

% ================= (a) Citeseer =================
\begin{tikzpicture}
\begin{axis}[
    width=6cm,
    height=4cm,
    xmode=log,
    log basis x=10,
    xmin=0.0002,
    xmax=2.5,
    ymin=27,
    ymax=72,
    xtick={0.001,0.01,0.1,1},
    xticklabels={
        $10^{-3}$,
        $10^{-2}$,
        $10^{-1}$,
        $10^{0}$
    },
    xlabel={Training time (sec)},
    ylabel={Accuracy (\%)},
    grid=both,
    grid style={gray!20},
    tick align=outside,
    tick pos=left,
    line width=0.7pt,
    label style={font=\scriptsize},
    tick label style={font=\scriptsize},
    clip=false
]

% Euclidean
\timepoint{blue}{+}{0.0011}{32.4}
\timepoint{red}{x}{0.0019}{38.1}
\timepoint{orange}{asterisk}{0.0668}{33.8}
\timepoint{green!60!black}{o}{0.0020}{44.2}
\timepoint{purple}{square}{0.0016}{39.8}
\timepoint{brown}{triangle}{0.0019}{41.4}
\timepoint{cyan!70!black}{diamond}{0.0025}{39.6}
\timepoint{magenta}{pentagon}{0.0072}{41.1}
\timepoint{teal!70!black}{square*}{0.0052}{32.9}
\timepoint{lime!70!black}{triangle*}{0.2876}{42.2}
\timepoint{gray}{diamond*}{0.0170}{51.7}
\timepoint{olive}{oplus}{0.0436}{53.2}
\timepoint{violet}{otimes}{0.0270}{34.0}
\timepoint{blue!40!black}{pentagon*}{0.1016}{53.0}
\timepoint{red!60!black}{*}{1.1478}{43.4}
\timepoint{green!30!black}{diamond}{0.0003}{35.1}
\timepoint{cyan!60!black}{diamond*}{0.0003}{57.7}
\timepoint{black}{square}{0.0003}{33.6}
\timepoint{orange!80!black}{triangle*}{0.0460}{57.0}

% Hyperbolic
\timepoint{pink!80!black}{x}{0.0146}{36.8}
\timepoint{blue!70!black}{oplus*}{0.0621}{57.3}
\timepoint{brown!70!black}{otimes*}{0.0259}{42.5}

% HyBridge: ★
\addplot[
    only marks,
    color=red,
    mark=filledstar,
    mark size=5pt,
    mark options={
        fill=red,
        draw=red
    }
] coordinates {(0.0480,60.7)};

\end{axis}
\end{tikzpicture}
&
% ================= (b) Cora =================
\begin{tikzpicture}
\begin{axis}[
    width=6cm,
    height=4cm,
    xmode=log,
    log basis x=10,
    xmin=0.0002,
    xmax=2.5,
    ymin=27,
    ymax=78,
    xtick={0.001,0.01,0.1,1},
    xticklabels={
        $10^{-3}$,
        $10^{-2}$,
        $10^{-1}$,
        $10^{0}$
    },
    xlabel={Training time (sec)},
    grid=both,
    grid style={gray!20},
    tick align=outside,
    tick pos=left,
    line width=0.7pt,
    label style={font=\scriptsize},
    tick label style={font=\scriptsize},
    clip=false
]

% Euclidean
\timepoint{blue}{+}{0.0010}{29.4}
\timepoint{red}{x}{0.0019}{45.1}
\timepoint{orange}{asterisk}{0.0918}{39.1}
\timepoint{green!60!black}{o}{0.0022}{52.5}
\timepoint{purple}{square}{0.0016}{43.1}
\timepoint{brown}{triangle}{0.0017}{45.5}
\timepoint{cyan!70!black}{diamond}{0.0027}{43.9}
\timepoint{magenta}{pentagon}{0.0075}{49.0}
\timepoint{teal!70!black}{square*}{0.0054}{29.3}
\timepoint{lime!70!black}{triangle*}{0.2813}{55.8}
\timepoint{gray}{diamond*}{0.0118}{61.0}
\timepoint{olive}{oplus}{0.0438}{55.8}
\timepoint{violet}{otimes}{0.0146}{37.2}
\timepoint{blue!40!black}{pentagon*}{0.1224}{61.5}
\timepoint{red!60!black}{*}{1.5526}{54.8}
\timepoint{green!30!black}{diamond}{0.0003}{37.0}
\timepoint{cyan!60!black}{diamond*}{0.0003}{61.9}
\timepoint{black}{square}{0.0003}{47.1}
\timepoint{orange!80!black}{triangle*}{0.0442}{62.5}

% Hyperbolic
\timepoint{pink!80!black}{x}{0.0117}{41.4}
\timepoint{blue!70!black}{oplus*}{0.0627}{57.2}
\timepoint{brown!70!black}{otimes*}{0.0258}{45.0}

% HyBridge: ★
\addplot[
    only marks,
    color=red,
    mark=filledstar,
    mark size=5pt,
    mark options={
        fill=red,
        draw=red
    }
] coordinates {(0.0452,67.2)};

\end{axis}
\end{tikzpicture}
\\[2mm]

\normalsize{(a) Citeseer}
&
\normalsize{(b) Cora}
\\

\end{tabular}%
}

% \vspace{-2mm}
\caption{The accuracy of HRL methods over the average training time per epoch (sec).}
\label{f13}
\vspace{-3mm}
\end{figure}

%% file: Figs/f11.tex
\begin{figure}[t]
\centering
\footnotesize

\begin{subfigure}[t]{0.48\linewidth}
\centering
\begin{tikzpicture}
\begin{axis}[
    width=5.3cm, height=2.4cm,
    xmin=0.1, xmax=1.0,
    ymin=0, ymax=3.0,
    xtick={0.2,0.4,0.6,0.8,1.0},
    xticklabels={0.2,0.4,0.6,0.8,1},
    ytick={0,1,2,3},
    grid=both,
    grid style={gray!20},
    tick align=outside,
    tick pos=left,
    line width=0.75pt,
    xticklabel style={font=\scriptsize},
    yticklabel style={font=\scriptsize},
    label style={font=\scriptsize}
]
\addplot[
    color=blue,
    thick,
    mark=*,
    mark size=1pt
] coordinates {
    (0.1,0.183045125)
    (0.2,0.4667537212)
    (0.3,0.6437799692)
    (0.4,0.9390553236)
    (0.5,1.183767605)
    (0.6,1.394758081)
    (0.7,1.678306651)
    (0.8,1.883869529)
    (0.9,2.196598029)
    (1.0,2.6260)
};
\end{axis}
\end{tikzpicture}
\vspace{-3mm}
\caption{\normalsize{DBLP-P}}
\end{subfigure}
\hfill
\begin{subfigure}[t]{0.48\linewidth}
\centering
\begin{tikzpicture}
\begin{axis}[
    width=5.3cm, height=2.4cm,
    xmin=0.1, xmax=1.0,
    ymin=0, ymax=1.5,
    xtick={0.2,0.4,0.6,0.8,1.0},
    xticklabels={0.2,0.4,0.6,0.8,1},
    ytick={0,0.5,1.0,1.5},
    grid=both,
    grid style={gray!20},
    tick align=outside,
    tick pos=left,
    line width=0.75pt,
    xticklabel style={font=\scriptsize},
    yticklabel style={font=\scriptsize},
    label style={font=\scriptsize}
]
\addplot[
    color=blue,
    thick,
    mark=*,
    mark size=1pt
] coordinates {
    (0.1,0.06463627815)
    (0.2,0.1120369196)
    (0.3,0.178905344)
    (0.4,0.2651489258)
    (0.5,0.3704082012)
    (0.6,0.4360704184)
    (0.7,0.4776086092)
    (0.8,0.5554839373)
    (0.9,0.6465022802)
    (1.0,0.8340)
};
\end{axis}
\end{tikzpicture}
\vspace{-3mm}
\caption{\normalsize{AMiner}}
\end{subfigure}

\caption{Training time per epoch of {\hbc} as the number of hyperedges increases.}
\label{f11}
\vspace{-1mm}
\end{figure}
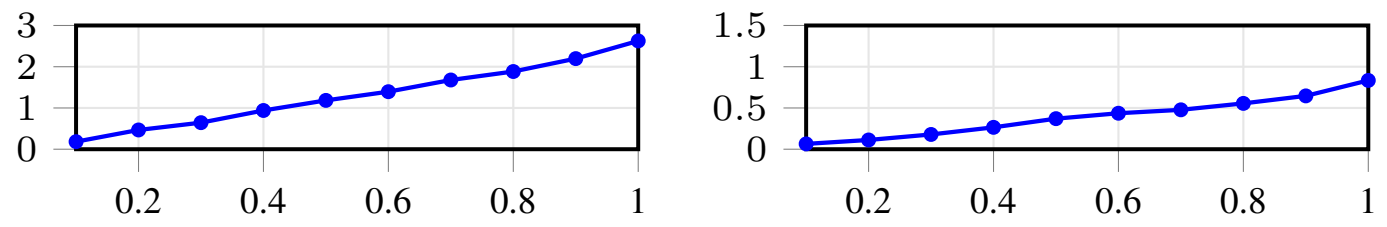

%% file: s6.tex
\section{Conclusions}\label{s6}
In this paper, we investigated the effectiveness of hyperbolic representation learning for hypergraphs and the structural guidance needed to effectively leverage hyperbolic space.
Our preliminary studies revealed that, although hypergraphs exhibit strong tree-like characteristics, simply extending existing HRL methods to hyperbolic space does not consistently improve downstream performance.
We further found that bridge scores provide effective structural guidance for organizing node representations along the radial dimension in graphs but do not directly generalize to hypergraphs due to their distinct groupwise connectivity, motivating {\hbc}, a self-supervised HB-HRL framework that estimates node bridge scores by explicitly accounting for the external connectivity of individual hyperedges and uses these scores to guide node radial positions together with contrastive learning.
Experiments on ten real-world hypergraphs against 22 competing methods demonstrated the effectiveness of {\hbc} across two downstream tasks (\ie, node classification and hyperedge prediction).
Further analyses showed that {\hbc} induces a more consistent center--periphery organization, incurs low preprocessing overhead, and exhibits favorable empirical scalability over the evaluated range.